\documentclass[aip,reprint]{revtex4-1}

\draft

\usepackage[utf8]{inputenc} 
\usepackage[english]{babel} 
\usepackage{amsmath}
\usepackage{amssymb}
\usepackage{enumitem} 
\usepackage{graphicx}
\usepackage{hyperref}
\usepackage{siunitx} 
\usepackage{xcolor} 
\usepackage{amsbsy}

\newcommand{\ee}{\mathrm{e}}
\newcommand{\ii}{\mathrm{i}}
\newcommand{\dd}{\,\mathrm{d}}
\newcommand{\w}{\omega}
\newcommand{\e}{\epsilon}
\def\pd2#1{\partial^2_{#1}}

\begin{document}

\title{Optimizing charge-balanced pulse stimulation for desynchronization} 

\author{Erik T. K. Mau}
\email[]{erikmau@uni-potsdam.de}
\affiliation{Department of Physics and Astronomy, University of Potsdam, 
Karl-Liebknecht-Str. 24/25, D-14476 Potsdam-Golm, Germany}

\author{Michael Rosenblum}
\email[]{mros@uni-potsdam.de}
\affiliation{Department of Physics and Astronomy, University of Potsdam, 
Karl-Liebknecht-Str. 24/25, D-14476 Potsdam-Golm, Germany}

\date{\today}

\begin{abstract}
Collective synchronization in a large population of self-sustained units appears both in natural and engineered systems.  Sometimes this effect is in demand, while in some cases, it is undesirable, which calls for control techniques. In this paper, we concentrate on pulsatile control, with the goal to either increase or decrease the level of synchrony. We quantify this level by the entropy of the phase distribution. Motivated by possible applications in neuroscience, we consider pulses of a realistic shape.  Exploiting the noisy Kuramoto-Winfree model, we search for the optimal pulse profile and the optimal stimulation phase. For this purpose, we derive an expression for the change of the phase distribution entropy due to the stimulus. We relate this change to the properties of individual units characterized by generally different natural frequencies and phase response curves and the population's state. We verify the general result by analyzing a two-frequency population model and demonstrating a good agreement of the theory and numerical simulations. 
\end{abstract}


\maketitle 

\begin{quotation}
Synchronization naturally emerges in interacting oscillatory systems, but often it is desirable to control its degree. A motivating example comes from neuroscience, where periodic pulse stimulation applied to the deep brain structures manipulates pathological rhythms in Parkinson's disease and other pathologies and reduces the symptoms. Hypothetically, this stimulation suppresses synchrony in a large population of coupled neurons.  This hypothesis triggered intensive research on control techniques that led to many feed-forward and closed-loop approaches to suppressing or enhancing synchrony. Potential applications in neuroscience imply additional requirements: the stimulation shall be pulsatile, and the pulses must be charge-balanced. It means the total current provided by the stimulus shall be zero to avoid the charge accumulation in the live tissue. This paper uses a paradigmatic model of coherent collective activity, namely the Kuramoto-Winfree model, to analyze the effect of charge-balanced pulse shape. We link the properties of individual units - their phase response curves - and the current state of the oscillatory ensemble to a collective response of the system to pulse stimulation. In this way, we optimize the stimulus shape and find the proper phase of the collective mode for the onset of stimulation. We support our theoretical findings by numerical simulation and argue that the validity of our results goes beyond the exploited simplistic model.
\end{quotation}

\section{\label{sec:intro} Introduction}

The emergence of a collective mode in a large oscillator 
network~\cite{Winfree-67,*Winfree-80,Kuramoto-75,*Kuramoto-84,*Strogatz-00,*Strogatz-03,*pikovsky2001} 
can be both beneficial and harmful. Examples of a positive role of collective synchrony include the coordinated firing of cardiac pacemaker cells~\cite{Winfree-67,*Winfree-80,Jalife-84} 
and coherent oscillation of interacting sources in a power grid~\cite{Doerfler-Bullo-12}. On the other hand, collective pedestrian synchrony on footbridges~\cite{Strogatz_et_al-05} represents a familiar example of an adverse effect of synchronization.  Further examples come from neuroscience, where neuronal populations' coherent activity manifests itself as a macroscopic brain rhythm~\cite{Breakspear-Heitmann-Daffertshofer-10}, whereas this rhythm can be either physiological or pathological. A particular example of an undesired rhythm is the enhanced brain activity in the beta band, i.e., between 13 and 30 Hz, which correlates to Parkinson's 
disease~\cite{little2013,*little2014,*tinkhauser2018}. The clinical treatment of the advanced stage of this pathology implies the administration of high-frequency electrical stimulation to a brain structure, typically to the subthalamic nucleus, via implanted microelectrodes~\cite{Benabid_et_al-91,*Benabid_et_al-09,*Kuehn-Volkmann-17}. This procedure is known as deep brain stimulation (DBS). Though the exact mechanism of DBS remains a matter of discussion, many theoretical and computational studies rely on the hypothesis formulated by P.A. Tass~\cite{Tass-99,*Tass-00,*Tass-01,*Tass_2001,*Tass-02} who treated the DBS task as a desynchronization problem and thus motivated research on the control of collective synchrony~\cite{Rosenblum-Pikovsky-04,*Rosenblum-Pikovsky-04a,%
Popovych-Hauptmann-Tass-05,%
Tukhlina-Rosenblum-Pikovsky-Kurths-07,Hauptmann-Tass-09,*Popovych-Tass-12,Montaseri_et_al-13,%
Lin_2013,*Zhou_2017,*Wilson-Moehlis-16,*Holt_et_al-16,Popovych_et_al-17,*Krylov-Dylov-Rosenblum-20,*Rosenblum-20,*Duchet_et_al-20}.  
In particular, numerous studies show that collective synchrony in a large, highly interconnected network can be efficiently suppressed or enhanced with the help of feedback control~\cite{Rosenblum-Pikovsky-04,*Rosenblum-Pikovsky-04a,%
Popovych-Hauptmann-Tass-05}.
This idea complies with the pilot experiments on closed-loop adaptive DBS~\cite{Rosin-11,*Little-13,*Cagnan_at_al-13,*Cagnan_at_al-17,*Holt1119,*McNamara2020.05.21.102335}.
However, not all control techniques meet the requirements of neuroscience applications, implying that stimulation has to be pulsatile and charge-balanced, as explained below. Motivated by the possible use of desynchronization techniques in DBS, in this paper, we address an essential aspect of this problem: optimizing the stimulus's shape.

The optimization problem was for the first time formulated by Wilson and Moehlis~\cite{wilson2014} who treated an ensemble of identical phase oscillators and applied calculus of variations to derive an optimal pulse's shape for a given oscillator's phase response curve (PRC). However, the obtained optimal waveforms are complex and, therefore, hardly implementable in practice. Indeed, commercially available DBS devices provide biphasic electrical stimuli, where each stimulus consists of two rectangular pulses of the opposite polarity~\cite{volkmann2002}. Below, we exploit the model of two interacting subpopulations of phase oscillators to analyze the effect of stimuli that are close in waveform to those produced by the standard DBS equipment. We examine the effect of the stimulus's shape; particularly, we investigate the impact of the time interval between the cathodic and anodic pulses.

The paper is organized as follows. In the rest of this Section, we discuss practical aspects and requirements 
of stimulation.  In Section \ref{sec:model} we introduce our model of nonidentical phase oscillators with frequency-dependent phase response curves and derive an expression for the macroscopic phase response.
In Section \ref{sec:information_entropy} we derive the formula for the variation of the phase distribution information entropy to be further used as a performance measure.
In Section \ref{sec:results} we use a particular case, a two-frequency phase oscillator model, to illustrate our theory and to verify it by numerics.
In Section \ref{sec:discussion} we summarize and discuss our results. The technical details are presented in Appendices. 

\subsection{\label{sec:intro_requirements} Requirements for neural tissue stimulation}
\begin{figure}
	\centering
	\includegraphics[width=0.48\textwidth]{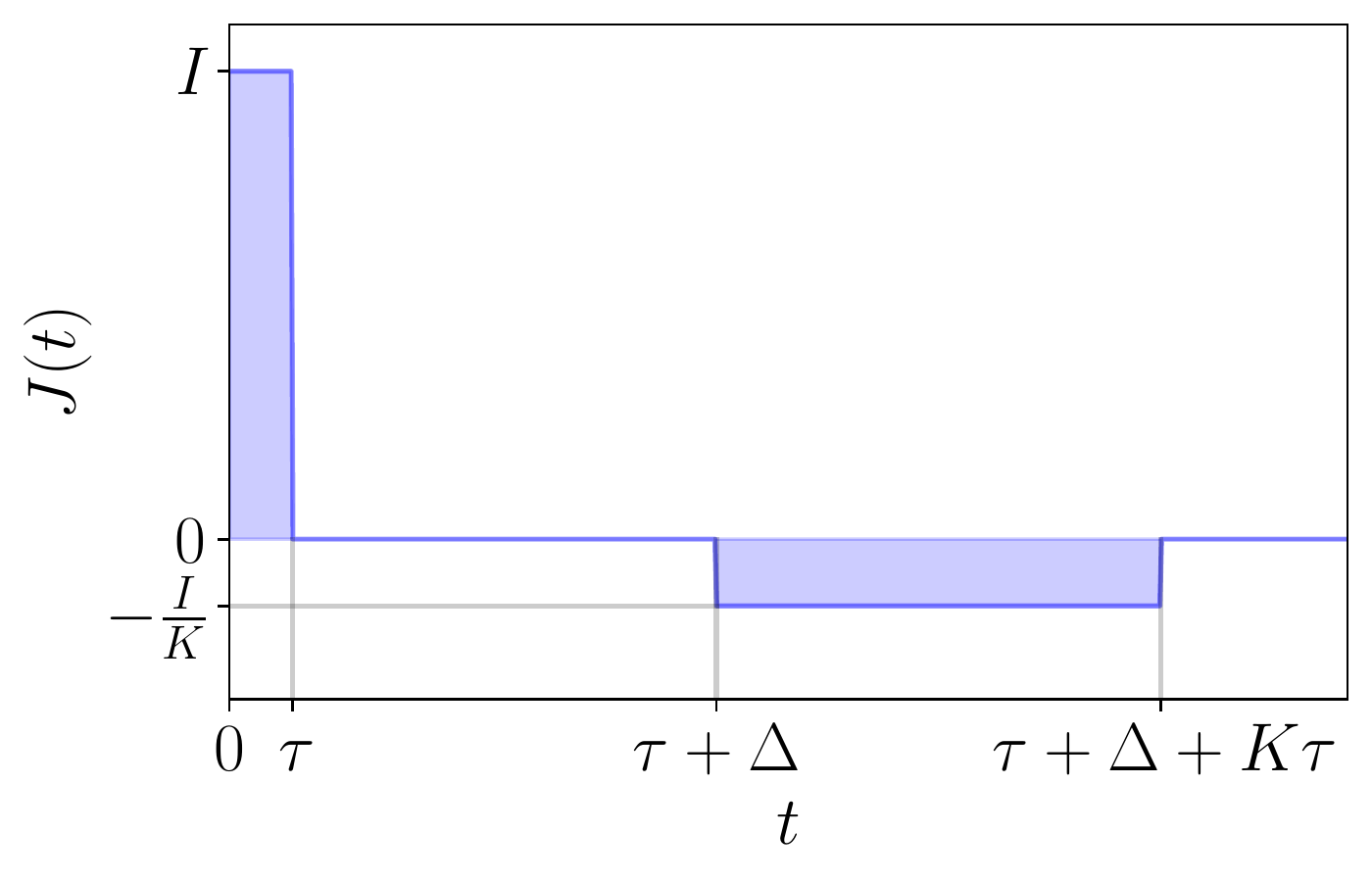}
    \caption{\label{fig:stimulus} Time dependence of the current for considered stimuli. Each stimulus consists of a rectangular pulse of amplitude $I$ and width $\tau$ and a following pulse of amplitude $-I/K$ and width $K\tau$; generally, $I$ can be negative. Notice that the parameters' choice ensures the fulfillment of the charge-balance condition Eq.~(\ref{eq:cb}).
The pulses are separated by a gap $\Delta$. The total stimulus's length is $T= \tau + \Delta + K\tau$.}
\label{fig:pulse}
\end{figure}

A typical DBS waveform is a sequence of constant-amplitude stimuli following with a frequency of about 130 Hz. 
A crucial safety requirement is to avoid tissue damage due to charge accumulation, which means that injected current $J(t)$ shall fulfill the condition  
\begin{equation}
    \int_0^T J(t)\dd t = 0 \;,
    \label{eq:cb}
\end{equation}
where $T$ is the total length of the stimulus. In other words, the stimulus has to be {\it charge-balanced}~\cite{merrill2005}.
Moreover, to exclude tissue damage by irreversible Faradaic reactions, the charge balance condition shall be fulfilled on a time scale $T\lesssim 1.5$ ms. 
Standard equipment provides a waveform consisting of two rectangular pulses, possibly with a small gap between them. In Fig.~\ref{fig:stimulus} we sketch the shape of the stimulus and introduce the notations for its four parameters.

In a most typical setting, the charge-compensating pulse immediately follows the first one, i.e., $\Delta=0$. However, stimulation with a gap between the pulses is possible~\cite{Butson-McIntyre-07}. Moreover, computational studies demonstrate that an increased gap can improve stimulation's efficiency~\cite{Foutz-McIntyre-10,*popovych2017a,*popovych2019}.
Here, we analyze the effect of the gap $\Delta$ and other parameters, $I,\tau,K$ on the stimulation efficiency.

\section{\label{sec:model} Population of noisy phase oscillators under charge-balanced stimulation}

This Section considers an effect of the charge-balanced rectangular stimulus sketched in Fig.~\ref{fig:pulse} 
on the synchrony level of an oscillator population.  Naturally, this effect depends on the stimulus's shape and the oscillation phase when the stimulus is applied. 

\subsection{\label{sec:model_1} The model}

We consider $N$ sine-coupled noisy phase oscillators
\begin{equation}
\dot{\phi}_i = \omega_i + \epsilon R \sin(\theta-\phi_i) + \xi_i(t)+J(t)Z(\phi_i; \omega_i) \;,
\label{eq:mod1}
\end{equation}
where $i=1,\ldots,N$, $\w_i$ are oscillators' frequencies, and $\e$ 
determines the coupling strength via the Kuramoto mean field
\begin{equation*}
   R \ee^{\ii\theta} = \frac{1}{N} \sum_{i=1}^N \ee^{\ii\phi_i} \;,
\end{equation*}
where $R$ and $\theta$ are field's amplitude and phase, respectively.
The oscillators are subject to zero-mean uncorrelated white noises, i.e.,
$\langle \xi_i(t) \rangle = 0$ and 
$\langle \xi_i(t) \xi_j(t') \rangle = 2D\delta_{ij} \delta(t-t')$, where $D$ is the diffusion coefficient. The last term in Eq.~(\ref{eq:mod1}) describes the oscillators' response to external stimulation. The stimulation $J(t)$ is common for all units, but there phase sensitivity functions (phase response curves, PRC)
$Z(\phi_i; \omega_i)$ are generally frequency-dependent.

In the thermodynamic limit $N \rightarrow \infty$ we describe the ensemble by the frequency distribution $g(\w)$ and the probability density function $P(\phi, t \,|\, \w)$ of oscillators with given natural frequency $\w$ at time instance $t$. The time evolution of $P$ is given by the partial differential equation (PDE)
\begin{align}
   \partial_t P
   = - \partial_\phi [ (\w+ \epsilon R \sin(\theta-\phi)
   + J Z) P ] 
   + D \pd2{\phi} P \;
   \label{eq:Kuramoto_PDE}
\end{align}
with the mean field defined as
\begin{align}
   R e^{\ii\theta}
   := \langle \ee^{\ii\phi} \rangle_{\phi}
   = \int_0^{2\pi} \ee^{\ii\phi} \rho(\phi,t) \dd\phi \,.
   \label{eq:mfield}
\end{align}
The global distribution $\rho(\phi,t)$ of oscillator's phases on the circle $[0,2\pi)$ at a given instance of time $t$ is the average of the conditional distribution $P$ over the natural frequencies:
\begin{equation}
  \rho(\phi,t) 
  =  \int_{-\infty}^\infty  P(\phi, t \,|\, \w) g(\w) \dd\w 
  \,.
  \label{eq:def_rho}
\end{equation}
Thus, the model combines the standard Kuramoto and Winfree terms and assumes that oscillators with equal frequencies $\w$ have the same PRC~\footnote{We emphasize, that the analysis and results below can be generalized to a case where the PRCs depend on one or more internal parameters $x$, not necessarily the natural frequencies as in the Kuramoto model. Then, the phase-conditional frequency-averaging $\langle \cdot \rangle_{\w|\phi}$ in the resulting expressions is replaced by the averaging over all internal parameters $\langle \cdot \rangle_{x|\phi}$ that the PRCs of the population depend on.}.
We note that a similar Kuramoto-Winfree model with identical PRC has already been used to explore the phase-dependence of effective DBS stimulation~\cite{weerasinghe2019}; however, the analysis in Ref.~\cite{weerasinghe2019} treated monophasic $\delta$-pulses only.

We analyze first the system in the absence of stimulation, $J(t)=0$. We assume that this autonomous oscillator population synchronizes, i.e., most of the oscillators form a smeared cluster that rotates around the unit circle with the frequency $\w_0$. In the following, we denote this state as the traveling wave (TW). In the TW state, the shape of the distributions $P(\phi, t \,|\, \w) = P(\phi-\omega_0 t, 0 \,|\, \w)$ is fixed and solely rotates with a constant frequency $\omega_0$, common for subpopulations with all natural frequencies $\omega$. This rotation is described by 
\begin{align}
   \partial_t P(\phi, t \,|\, \w)
   = - \omega_0 \partial_\phi P(\phi, t \,| \w)  
   \qquad \forall~\w 
   \;.
   \label{eq:rotate}
\end{align}
Thus, the time evolution of the mean-field phase angle $\theta$ is given by $\theta = \omega_0 t + \theta(0) = \w_0t + \theta_0$ and can be used to unambiguously determine the position of $P$ by
\begin{align}
    P(\phi,t \,|\, \w) = P(\phi - \theta, t_0 \,|\, \w) = P_\text{\tiny TW}(\phi-\theta \,|\, \w)
    \,,
\end{align}
where $t_0 := -\theta_0/\omega_0$ is a time point where $\theta=0$. We use this time point to fix the distributions $P_\text{\tiny TW}(\phi \,|\, \w)$ to have the shape of the $\w$-dependent TW solutions and a position where $\theta=0$. In the same manner we define $\rho_\text{\tiny TW}$. By plugging Eq.~(\ref{eq:rotate}) into Eq.~(\ref{eq:Kuramoto_PDE}), we obtain, for each frequency $\w$, 
\begin{align}
   0 = - \partial_\phi [ (\w - \w_0 - \epsilon R \sin(\phi))P_\text{\tiny TW}] 
   + D \pd2{\phi} P_\text{\tiny TW} 
   \;.
    \label{eq:TW_ODE}
\end{align}
These equations define the shape of the TW solutions $P_\text{\tiny TW}(\phi \,|\,\w)$. The time dependence of phase averages evaluated at the TW state can therefore be formulated as functions of the mean-field phase $\theta$ by
\begin{align}
    \langle f \rangle_\phi 
    =& \int_0^{2\pi} f(\phi) \rho(\phi,t) \dd\phi \nonumber \\
    =& \int_0^{2\pi} f(\phi) \rho_\text{\tiny TW}(\phi-\theta) \dd\phi \nonumber \\
    =& \int_0^{2\pi} f(\phi + \theta) \rho_\text{\tiny TW}(\phi) \dd\phi \nonumber \\
    =& \langle f(\phi+\theta) \rangle_\phi^\text{\tiny TW} 
    \label{eq:average_TW}
    \,,
\end{align}
exploiting the $2\pi$-periodicity in $\phi$. We denote the evaluation of phase averages at the TW state with the superscript 'TW'. Analogously, we denote the conditional phase average $\langle \cdot \rangle_{\phi|\w}$ that is evaluated at the TW state by $\langle \cdot \rangle^\text{\tiny TW}_{\phi|\w}$.

\subsection{\label{sec:model_phase_osc} Uncoupled identical oscillators}
First, following Wilson and Moehlis\cite{wilson2014}, we consider the simplest case of identical noisy oscillators with the same PRC $Z$. Coupling tends to make their phases equal, but noise counteracts the synchronizing action. Thus, we can assume $\phi_i\approx\theta$.
Let us choose two oscillators with phases $\phi_{k,j}=\theta+\psi_{k,j}$,  where $|\psi_{k,j}|\ll 1$, and look for the evolution of the phase difference $\phi_k-\phi_j=\psi_k-\psi_j=\psi$ under effect of stimulation. 
We neglect the effects of coupling and noise for the duration of the stimulus. Then, the dynamics of oscillators obeys: 
\[
\dot\phi_{k,j}=\w+J(t)Z(\phi_{k,j})=
\w+J(t)Z(\theta+\psi_{k,j})\;.
\]
Using smallness of $\psi_{k,j}$, we write $Z(\phi_{k,j})=Z(\theta)+ Z'\psi_{k,j}$ and obtain $\dot\psi=J(t)Z' \psi$. 

Let the time of the stimulus's application be $t=0$ and let us denote 
$\psi(0)=\psi_0$,  $\psi(\tau)=\psi_1$, $\psi(\tau+\Delta)=\psi_2$,  $\psi(T)=\psi_3$, and similarly for $\theta$. Then, after the first pulse we have:
\begin{align}
    \ln\psi_1-\ln\psi_0=\int_0^\tau J Z'(\theta(t))\dd t
    \approx \tau I Z'(\theta_0)\;.
    \label{eq:phase_osc_mono}
\end{align}
Similarly, the evolution of the phase difference
during the second pulse is
\[
\ln\psi_3-\ln\psi_2
\approx -\tau I Z'(\theta_2)\;.
\]
However, $\psi$ does not vary within the gap $\Delta$, i.e., $\psi_1=\psi_2$, and hence
\begin{align}
    \ln(\psi_3/\psi_0)
    &=-I\tau\left[Z'(\theta_2)-Z'(\theta_0)\right ]
    \;,
\end{align}
which, for small $\Delta$, can be approximated by
\begin{align}
    \ln(\psi_3/\psi_0)
     &\approx
    -I\tau Z''(\theta_0)(\tau+\Delta)
    \;.
    \label{eq:phase_osc_bi}
\end{align}
Thus, to maximize the desynchronizing effect, we have to stimulate around the mean-field phase $\theta_0$ corresponding to the maximum of $|Z''(\theta)|$; the sign of $I$ shall be opposite to the sign of $Z''$. 
We see that this simple approximation exhibits a dependence on the gap length $\Delta$. 

\subsection{\label{sec:model_macroresponse} From microscopic PRC to macroscopic response}
Now, we return to the general case of non-identical oscillators in a TW state. Although single oscillators can drift asynchronously, the opposite effects of coupling and diffusion form a partially synchronous state with nonzero constant mean-field amplitude $R$. 
For the subpopulation $P(\phi, t \,|\, \w)$ with the natural frequency $\w$, we describe the evolution during the stimulation as
\begin{align}
   \partial_t P   
   = - \partial_\phi  \left[ \left(
   \omega_0 + J(t) Z(\phi;\w)
    \right) P   \right] 
    \;.
    \label{eq:TW_PDE_w}
\end{align}
This expression is exact only at the onset of stimulation at $t=0$. By assuming the validity of this PDE for the entire time of the stimulation $0 \leq t \leq T$, when the system is perturbed from the TW state, we make two assumptions: First, we assume $I$ to be larger than the effect of coupling and diffusion, that stabilizes the TW state. Secondly, we assume $2\pi/\omega_0$ to be larger than the relaxation time. In this way, we can neglect the terms of coupling and diffusion during the two stimulation intervals and the off-stimulation time.

Averaging over the frequency distribution $g(\w)$ yields the time evolution of the global phase distribution $\rho$:
\begin{align}
   \partial_t \rho(\phi,t) 
   = - \partial_\phi \left[ \left( 
   \omega_0 
   + J(t) \langle Z \rangle_{\omega|\phi}
    \right) \rho(\phi,t) \right ] 
    \;,
    \label{eq:TW_PDE}
\end{align}
where we made use of Bayes' theorem about conditional probabilities to express the conditional frequency distribution by
\begin{align}
    P(\w,t \,|\, \phi) = \frac{P(\phi,t \,|\, \w)g(\w)}{\rho(\phi,t)}
    \label{eq:bayes_theorem}
\end{align}
and to obtain the conditional frequency average of $Z$:
\begin{align}
    \langle Z \rangle_{\omega|\phi} 
    =&~ \int_{-\infty}^\infty  Z(\phi; \omega) P(\w,t \,|\, \phi) \dd\w 
    \;.
\label{eq:mod2}
\end{align}
Here, $P(\w,t \,|\, \phi) \dd \w$ is the probability to find an oscillator with natural frequency in the range $[\omega, \omega + \dd \w)$ given that its position is $\phi$ at time instance $t$. Due to the time dependence of the distributions $\rho$, $P(\phi \,|\, \w)$, and $P(\phi \,|\, \w)$, averaging a static function, e.g., $Z$, with respect to one of them adds a time-dependence. A summary of the notation of statistical averages and their relations can be found in Appendix~\ref{sec:apx_stat_averages}.

To show the relationship between the PRCs as characteristics of the microscopic stimulation response of single oscillators and the macroscopic response of the mean-field phase angle $\theta$, we derive its time evolution equation in the TW state, that is perturbed by stimulation. Differentiating Eq.~(\ref{eq:mfield}) with respect to time we obtain
\begin{align}
    \dot{\theta} 
    = \frac{1}{R}\Im \left( e^{-i\theta} \frac{\dd}{\dd t} \langle \ee^{\ii \phi} \rangle_{\phi} \right)
    \,,
\end{align}
where $\Im$ denotes the imaginary part. Using the expression for $\partial_t\rho$ from Eq.~(\ref{eq:TW_PDE}), we write
\begin{align}
    \frac{\dd}{\dd t} \langle \ee^{\ii \phi} \rangle_{\phi}
    = \ii \omega_0 R\ee^{\ii \theta} + \ii J(t) \langle \ee^{\ii \phi} Z \rangle_{\phi, \w}
\end{align}
and thus
\begin{align}
	\dot{\theta} 
	=& \w_0 + \frac{J(t)}{R} \langle Z \cos(\theta - \phi) \rangle_{\phi, \w}\;.
\end{align}
Note, that in the TW state the time-dependence which comes automatically with averaging, can be encoded with $\theta$ itself, as outlined in Sec.~\ref{sec:model_1}. We define the macroscopic PRC for the phase response of the mean field 
\begin{align}
    \mathcal{Z}(\theta) = \frac{1}{R} \langle Z(\phi + \theta) \cos(\phi) \rangle^\text{\tiny TW}_{\phi, \w}\;,
\end{align}
such that, with the assumption that the stimulation does not change the shape of $\rho$ significantly, the ODE for $\theta$ can be formulated in the Winfree form as
\begin{align}
    \dot{\theta} 
	=& \w_0 + J(t) \mathcal{Z}(\theta)
	\,.
	\label{eq:ODE_theta}
\end{align}
$\mathcal{Z}$ as a function of $\theta$ can be calculated from the PRCs $Z(\phi; \omega)$ and the shape of the TW state $P_\text{\tiny TW}(\phi \,|\, \w)$.

\section{\label{sec:information_entropy} Performance measure: the information entropy}

The mean-field amplitude $R$ does not provide a complete characterization of synchrony. Indeed, it vanishes both in the completely incoherent state (IC) with $\rho=(2\pi)^{-1}$ and, e.g., in the two-cluster state. Therefore, we exploit the information entropy~\cite{Tass_et_al-98}
\begin{align}
   H(t)= - \int_0^{2\pi} \rho(\phi,t) \ln[\rho(\phi,t)] \dd \phi
   \label{eq:H_def}
\end{align}
as a universal measure of incoherence since only the IC state yields maximum entropy $H$. Hence, we use the variation of $H$ due to the stimulus as the performance measure of the stimulation. A total negative change means a step towards synchronization whereas a positive total change in entropy means a step towards desynchronization.

To take into account the invasiveness of the stimulation, we normalize the entropy change by the charge
$\int_0^\tau J(t) \,\dd t = I\tau$ 
injected by the first pulse~\footnote{We remind that for a charge-balanced stimulation the total current $\int_0^T J(t) \,\dd t =0$. The quantity $\int_0^T |J(t)| \,\dd t = 2 |I|\tau$ describes how invasive the stimulus is.}.
In this way, we define
\begin{align}
    h(t) = \frac{H(t) - H(0)}{I\tau}
    \label{eq:h_def}
\end{align}
as the total entropy change per injected charge. 

At first, we derive an $\mathcal{O}(\tau)$-approximation for $h$ after monophasic stimulation, i.e. after the first pulse of the biphasic stimulus. We assume that (i) the system is in a TW state for the duration of the stimulation $\tau$, and thus obeys Eqs.~(\ref{eq:TW_PDE_w},\ref{eq:TW_PDE}) for $0 \leq t \leq \tau$ and (ii) the stimulation is short compared to the oscillation period, i.e. $\tau \ll 2\pi/\w_0$. Then, $h$ can be approximated in the first order of $\tau$ by
\begin{align}
   h(\tau)
   \approx \frac{\dot{H}(0)}{I} 
   = \langle \partial_\phi \langle Z(\phi + \theta_0) \rangle^\text{\tiny TW}_{\w|\phi} \rangle^\text{\tiny TW}_\phi
   \;.
   \label{eq:H_mono_approx}
\end{align}
The derivation of the entropy time derivative $\dot{H}$ can be found in detail in Appendix~\ref{sec:apx_Hder}. 

Consider now a state $\theta^-$ that minimizes the PRC-dependent right hand side in Eq.~(\ref{eq:H_mono_approx}): 
\begin{align}
    h_\text{min}=&\langle \partial_\phi \langle Z(\phi + \theta^- \rangle^\text{\tiny TW}_{\w|\phi} \rangle^\text{\tiny TW}_\phi \label{eq:h_min} \\ \nonumber
    =& \min_{\theta_0 \in [0,2\pi)} \langle \partial_\phi \langle Z(\phi + \theta_0 \rangle^\text{\tiny TW}_{\w|\phi} \rangle^\text{\tiny TW}_\phi \;.
\end{align}
In the same way, we define the state $\theta^+$ that yields $h_\text{max}$~\footnote{Generally the definition of $\theta^\mp$ can be ambiguous if several values of $\theta$ provide the same extreme value.}. For a positive mono-pulse, $I>0$, $\theta^\mp$ are the optimal phases for the synchronizing and desynchronizing action, respectively~\footnote{Note, that if $h_\text{min}>0$, then  both actions decrease the synchrony level.}. For $I<0$, the stimulation effect is reversed.

We emphasize the analogy to the phase response of the phase oscillator model to a single pulse given by Eq.~(\ref{eq:phase_osc_mono}). In that case, the first derivative of the PRC determines the optimal phase for the (de)synchronizing stimulation. For our more general model, it is the phase average of the first derivative of the conditional frequency-averaged PRC instead of simply the first derivative of the PRC. 

The next step is to derive an approximation for the entropy change after the entire biphasic stimulus.
We again assume that the system is in a TW state for the entire duration of the stimulation $T$ and that both stimulation intervals are short if compared to the oscillation period, $\tau \ll 2\pi/\w_0$ and $K\tau \ll 2\pi/\w_0$. We split the stimulus's duration $T$ into three intervals, due to the discontinuity of $J(t)$ at $t=\tau$ and $t=\tau + \Delta$ and write
\begin{equation}
H(T) - H(0) \approx \tau \dot{H}(0) + \Delta \dot H(\tau)
    + K\tau \dot{H}(\tau + \Delta) \;.
\label{eq:H_step_biphasic}
\end{equation}
$\dot H(0)$ is given by Eq.~(\ref{eq:H_mono_approx}). Applying this equation to other time intervals, we obtain $\dot H(\tau)=0$ and  
\begin{align}
    \dot{H}(\tau + \Delta) = -\frac{I}{K} \langle \partial_\phi \langle Z(\phi + \theta(\tau+\Delta)) \rangle^\text{\tiny TW}_{\w|\phi} \rangle^\text{\tiny TW}_\phi
    \label{eq:dH_tau+Delta}
    \,.
\end{align}
Using the TW approximation we write $\theta(\tau + \Delta) \approx \theta_0 + \w_0 (\tau + \Delta)$ and thus $Z$ can be written in $\mathcal{O}(\tau)$ approximation as
\begin{align}
    Z(\phi + \theta(\tau + \Delta))
    &\approx Z(\phi + \theta_0 + \w_0\Delta) \nonumber \\
    &+ \w_0 \tau \partial_\phi Z(\phi + \theta_0 + \w_0\Delta)
    \label{eq:Z_tau+Delta}
    \,,
\end{align}
where we omit the $\w$-dependence in $Z$ for clarity of notation.
Inserting Eq.~(\ref{eq:Z_tau+Delta}) into Eq.~(\ref{eq:dH_tau+Delta}), we obtain an approximate expression for $\dot{H}(\tau+\Delta)$. Substituting 
$\dot{H}(\tau+\Delta)$ and $\dot H(0)$ into  Eq.~(\ref{eq:H_step_biphasic}) and using Eq.~(\ref{eq:h_def}) we obtain the  entropy change per injected charge:
\begin{align}
   h(T) 
   &\approx
   \langle \partial_\phi \langle Z(\phi+\theta_0) - Z(\phi + \theta_0 + \w_0 \Delta) \rangle^\text{\tiny TW}_{\w|\phi} \rangle^\text{\tiny TW}_\phi \nonumber \\ 
   &~- \w_0 \tau \langle \partial_\phi \langle \partial_\phi Z(\phi + \theta_0 + \w_0\Delta) \rangle^\text{\tiny TW}_{\w|\phi} \rangle^\text{\tiny TW}_\phi
    \;.
\end{align}
Note that in the $\mathcal{O}(\tau)$-approximation, the dependence on $K$ cancels out. 

Next, we consider two limit cases. For short off-stimulation times $\w_0 \Delta \ll 1$, we keep only the terms of the first order in $\tau$ and $\Delta$ and obtain:
\begin{align}
   h(T)
   \approx - \w_0 (\tau + \Delta) 
   \langle \partial_\phi \langle \partial_\phi Z(\phi+\theta_0) \rangle^\text{\tiny TW}_{\w|\phi} \rangle^\text{\tiny TW}_\phi
   \;.
   \label{eq:H_bi_approx_Delta_small}
\end{align}
According to this expression, the stimulation's effect is the strongest at the initial state $\theta_0 = \theta_\text{bi}$ that maximizes
\begin{align}
    \left|\langle \partial_\phi \langle \partial_\phi Z(\phi+\theta_0) \rangle^\text{\tiny TW}_{\w|\phi} \rangle^\text{\tiny TW}_\phi \right|
    \label{eq:theta_bi}
    \,.
\end{align}
(For the same reason as for the mono-pulse stimulation, the definition of $\theta_\text{bi}$ can be ambiguous.) The phase $\theta_\text{bi}$ is the most sensitive to biphasic stimulation in the limit of small $\tau$ and $\Delta$. However, the stimulation at $\theta_\text{bi}$ can either increase or decrease the entropy, depending on the sign of $I$. For a desynchronizing effect, the sign of $I$ shall be opposite to $\langle \partial_\phi \langle \partial_\phi Z(\phi + \theta_0) \rangle^\text{\tiny TW}_{\w|\phi} \rangle^\text{\tiny TW}_\phi$. 
There is a clear analogy between this result and that for uncoupled identical oscillators, cf. Eq.~(\ref{eq:phase_osc_bi}). The latter can be recovered in the case of identical oscillators and a Dirac's-$\delta$ phase distribution.

In the second limit case of larger $\Delta$ comparable to the oscillation period, i.e. $\w_0 \Delta \approx 1$, the dominating term
\begin{align}
   h(T)
   \approx
   \langle \partial_\phi \langle Z(\phi+\theta_0) - Z(\phi + \theta_0 +   \w_0 \Delta) \rangle^\text{\tiny TW}_{\w|\phi} \rangle^\text{\tiny TW}_\phi 
    \label{eq:H_bi_approx_Delta_large}
\end{align}
represents an averaged difference between the PRC and its shifted version. In the simplest approximation, we assume an additive effect of two consecutive pulses and no entropy change between them. Then we design the most efficient stimulation choosing $\theta_0=\theta^-$, 
$\theta_0+\w_0\Delta=\theta^+$, for $I<0$, 
or $\theta_0=\theta^+$, 
$\theta_0+\w_0\Delta=\theta^-$, for $I>0$.
The corresponding entropy change per injected charge is
\begin{align}
    h(T) = 
    \langle \partial_\phi \langle Z(\phi + \theta^\mp) 
    - Z(\phi + \theta^\pm) \rangle^\text{\tiny TW}_{\w|\phi} \rangle^\text{\tiny TW}_\phi  \;.
    \label{eq:h_max}
\end{align}
Notice that in this approximation $h(T)=h_\text{min}-h_\text{max}$ for $I<0$ and $h(T)=h_\text{max}-h_\text{min}$ for $I>0$. In the following we compare this approximation with the actual entropy changes due to biphasic stimulation.

\section{Verification of the theory}
\label{sec:results}

We test the derived approximate expression for the total entropy change after both mono- and biphasic pulsatile stimulation from Sec.~\ref{sec:model}, using a particular example of a two-group Kuramoto-Winfree model. The oscillators within each group are identical, and each group has its own frequency. First, we outline the theoretical results for this special case and then compare them to numerical results in Sec.~\ref{sec:results_mono} and \ref{sec:results_bi}. 

\begin{figure*}
	\centering
	\includegraphics[width=0.9\textwidth]{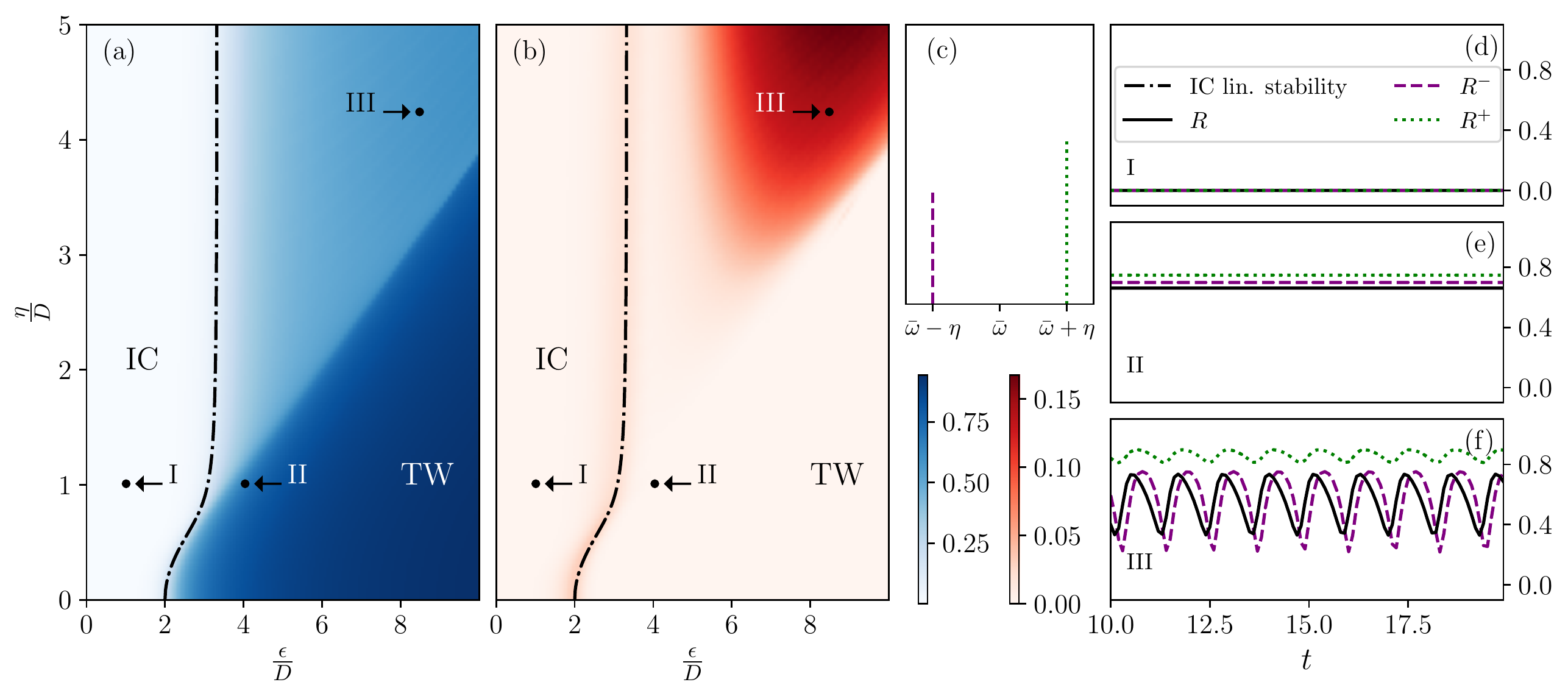}
    \caption{\label{fig:bifurcation_diagram} Dynamical regimes of bimodal Kuramoto model for $\alpha=0.4$: Bifurcation diagrams (a) and (b) show in color code the value of time-averaged mean value (blue) and standard deviation (red) of the mean field amplitude $R(t)$ in equilibrium, respectively. For all parameter combinations of $\epsilon$ and $\eta$, the system starts at the same initial state that is specified in Appendix~\ref{sec:apx_methods}. Time averages are taken after a transient time of $t_\text{eq} = 8$ for an averaging period of $T_\text{av} = 10$. Three qualitatively different equilibrium states are revealed: The incoherent states (IC), with example in (d) ($\frac{\epsilon}{D}=1$, $\frac{\eta}{D}=1$), the traveling-wave states (TW), with example in (e) ($\frac{\epsilon}{D}=4$, $\frac{\eta}{D}=1$), and the region of oscillating $R(t)$, with example in (f) ($\frac{\epsilon}{D}=8.5$, $\frac{\eta}{D}=4.2$). The time evolution plots (d,e,f) show the mean field amplitudes $R^\mp$ and $R$ and are obtained using the system parameters at points marked by I, II, and III in (a,b). The boundary of linear stability for the incoherent states \cite{acebron1998} is marked by the black solid line. Due to the multi-stability in regions close to the transitions between different regimes \cite{acebron1998}, the bifurcation diagram can differ for other initial conditions. (c): Depiction of bimodal frequency distribution $g$: The slow (violet) and fast (green) population are marked as columns with height $\alpha$ and $1-\alpha$, respectively. For all numerical simulations, $D=1$ and $\bar{\omega}=10$ are chosen.
    }
\end{figure*}

\subsection{\label{sec:results_bimodal} A particular case: the two-frequency ensemble}

To make the first step towards a realistic model that should include excitatory and inhibitory neuronal subpopulations, we choose $g(\w)$ to be a sum of two delta-functions, cf.~\cite{acebron1998}:
\begin{align}
     g(\w) = \alpha \delta(\w - \w^-)
+ (1-\alpha) \delta(\w - \w^+)\;,
    \label{eq:g_bimodal}
\end{align}
where $\w^\mp$ are the frequencies of the slow and fast subpopulations. 
In the following, we will repeatedly use the superscripts $\mp$ for quantities belonging to one of the two subpopulations. They shall be distinguished from same superscripts used in Sec.~\ref{sec:information_entropy}. The mean-field amplitudes of both subpopulations are denoted by
\begin{align}
    R^\mp 
    = |\langle \ee^{\ii \phi} \rangle_{\phi | \w^\mp}|
    = \left| 
    \int_0^{2\pi}  \ee^{\ii \phi} P(\phi,t \,|\, \w^\mp) \dd\phi 
    \right|
    \;,
\end{align}
whereas the global mean field is given by
\begin{align}
    Re^{i\theta} 
    = \langle \ee^{\ii \phi} \rangle_{\phi} 
    = 
    \alpha \langle \ee^{\ii \phi} \rangle_{\phi | \w^-} + (1-\alpha) \langle \ee^{\ii \phi} \rangle_{\phi | \w^+}
    \;.
\end{align}
For the analysis of the unperturbed system, it is convenient to set $\w^\mp=\Bar{\w}\mp \eta$ with the central frequency $\bar{\w}$ and the frequency detuning $\eta$. We illustrate the ensemble's dynamical regimes in Fig.~\ref{fig:bifurcation_diagram} for the parameters 
$\alpha=0.4$, $D=1$, and $\bar\w=10$. Although the landscape of dynamical regimes depends only on $\alpha$, $\frac{\epsilon}{D}$ and $\frac{\eta}{D}$, see~\cite{acebron1998}, the values of $\bar{\w}$ and $D$ become important when stimulation is introduced, as their ratio determines the rate of relaxation to the stable state.

In the following, we proceed with these parameters and additionally choose the TW scenario described in Fig.~\ref{fig:bifurcation_diagram} as the equilibrium state for the stimulation (see the symbols marked with II in panels (a,b)). Thus, we fix the coupling parameter $\epsilon=4$ and the frequency detuning $\eta = 1$. In this TW state, both subpopulations rotate with some common frequency $\w_0 \in [\bar{\w}-\eta, \bar{\w}+\eta]$. As described in Sec.~\ref{sec:model_1}, the time-dependence of the system is then entirely captured by the mean-field phase $\theta$.

\begin{figure}
    \centering
    \includegraphics[width=0.48\textwidth]{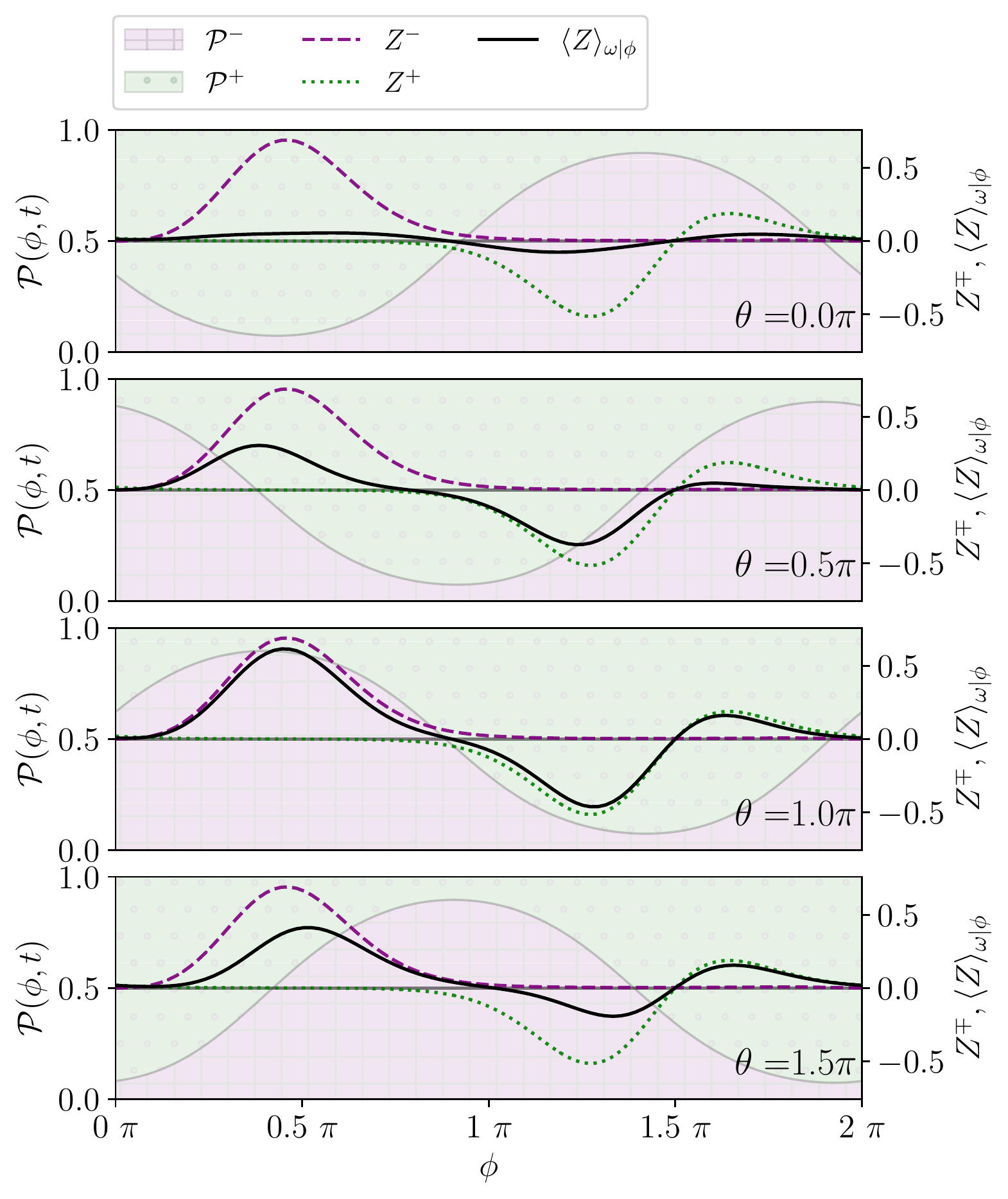}
    \caption{Bimodal Kuramoto-Winfree model in a TW state with system parameters $\bar{\w} = 10$, $\eta=1$, $D=1$, $\epsilon=4$, $\alpha=0.4$. Left $y$-axis: Conditional probabilities $\mathcal{P}$ as functions of $\phi$ for four different states $\theta$. At any $\phi$ the ratio of magenta- and green-filled space denotes the conditional probabilities $\mathcal{P}^-(\phi)$ and $\mathcal{P}^+(\phi)$, respectively. 
    Right y-axis: Static PRCs $Z^\mp$ (dotted and dashed curves) and $\theta$-dependent $\langle Z \rangle_{\omega|\phi}$ (solid curve) as functions of $\phi$.
     }
    \label{fig:cond_freq_av_PRC}
\end{figure}

We denote the conditional probabilities to find an oscillator of frequency $\omega^\mp$ given its phase $\phi$ at time $t$ (or at mean-field phase $\theta$) by
\begin{align}
    \mathcal{P}^\mp(\phi, t) 
    = \lim_{\gamma \rightarrow 0} \int_{\w^\mp-\gamma}^{\w^\mp+\gamma} 
    P(\w, t \,|\, \phi) \dd\w 
\end{align}
and calculate them integrating over the conditional frequency distribution in the vicinity of $\w^\mp$. Using Eqs.~(\ref{eq:bayes_theorem},\ref{eq:g_bimodal}) we obtain
\begin{align}
    \mathcal{P}^\mp(\phi, t) = 
    (0.5 \mp (0.5-\alpha))
    \frac{P(\phi,t \,|\, \w^\mp)}{\rho(\phi,t)} 
    \;,
\end{align}
where the global phase distribution $\rho$ is calculated via
\begin{align}
    \rho(\phi,t) = \alpha P(\phi,t \,|\, \w^-) + (1-\alpha) P(\phi,t \,|\, \w^+)
    \label{eq:global_phase_distr_bimodal}
    \,.
\end{align}
Normalization of probabilities ensures $\mathcal{P}^{-}(\phi,t) + \mathcal{P}^{+}(\phi,t) = 1$. Figure~\ref{fig:cond_freq_av_PRC} shows $\mathcal{P}^\mp$ as functions of $\phi$ for different states $\theta$.

The PRCs of subpopulations with frequencies $\omega^\mp$ are denoted as $Z^\mp(\phi)$, respectively. PRC of neuronal oscillators are typically classified 
as types I and II~\cite{Hansel-Mato-Meunier-95,Canavier-06}. 
For the slow subpopulation we choose a type-I PRC modeled by~\cite{cestnik2017}
\begin{align}
    Z^-(\phi) = (1-\cos(\phi))\exp{[3(\cos(\phi-\pi/3)-1)]}
    \;.
\end{align}
PRCs of type-I take only positive values, hence a stimulation with positive current $I$ always yields an advance in phase, whereas a stimulation with negative current $I$ leads to a delay in phase. On the contrary, the fast population is chosen to have a type-II PRC modeled by~\cite{cestnik2017} 
\begin{align}
    Z^+(\phi) = \cos(\phi)\exp{[3(\cos(\phi-1.4\pi)-1)]}
    \;.
\end{align}
A type-II PRC can take both negative and positive values, hence the positive stimulation $I>0$ can advance or delay the phase, depending on the system's state.

The conditional frequency-averaged PRC defined in Eq.~(\ref{eq:mod2}) can thus be expressed through $Z^\mp$ and $\mathcal{P}^\mp$ as 
\begin{align}
    \langle Z \rangle_{\omega|\phi} = 
   Z^{-}(\phi) \mathcal{P}^-(\phi,t) + Z^{+}(\phi) \mathcal{P}^+(\phi,t)
    \;.
\end{align}
This $\theta$-dependent function as well as the common PRCs $Z^\mp$ are depicted in Fig.~\ref{fig:cond_freq_av_PRC}. Obviously, $\langle Z \rangle_{\omega|\phi}$ can only attain values between the two static curves $Z^\mp$. It is large at those states $\theta$ where a large part of one population attains some position $\phi$ for which its PRC is also significant.  Thus, $\langle Z \rangle_{\omega|\phi}$ is close to zero at $\theta = 0$ and almost maximal at $\theta=\pi$ in Fig.~\ref{fig:cond_freq_av_PRC}. In the following, we use this setup to study the ensemble's entropy change in response to mono- and biphasic stimuli.

\subsection{\label{sec:results_mono} Response to monophasic stimulation}

\begin{figure}
	\includegraphics[width=0.48\textwidth]{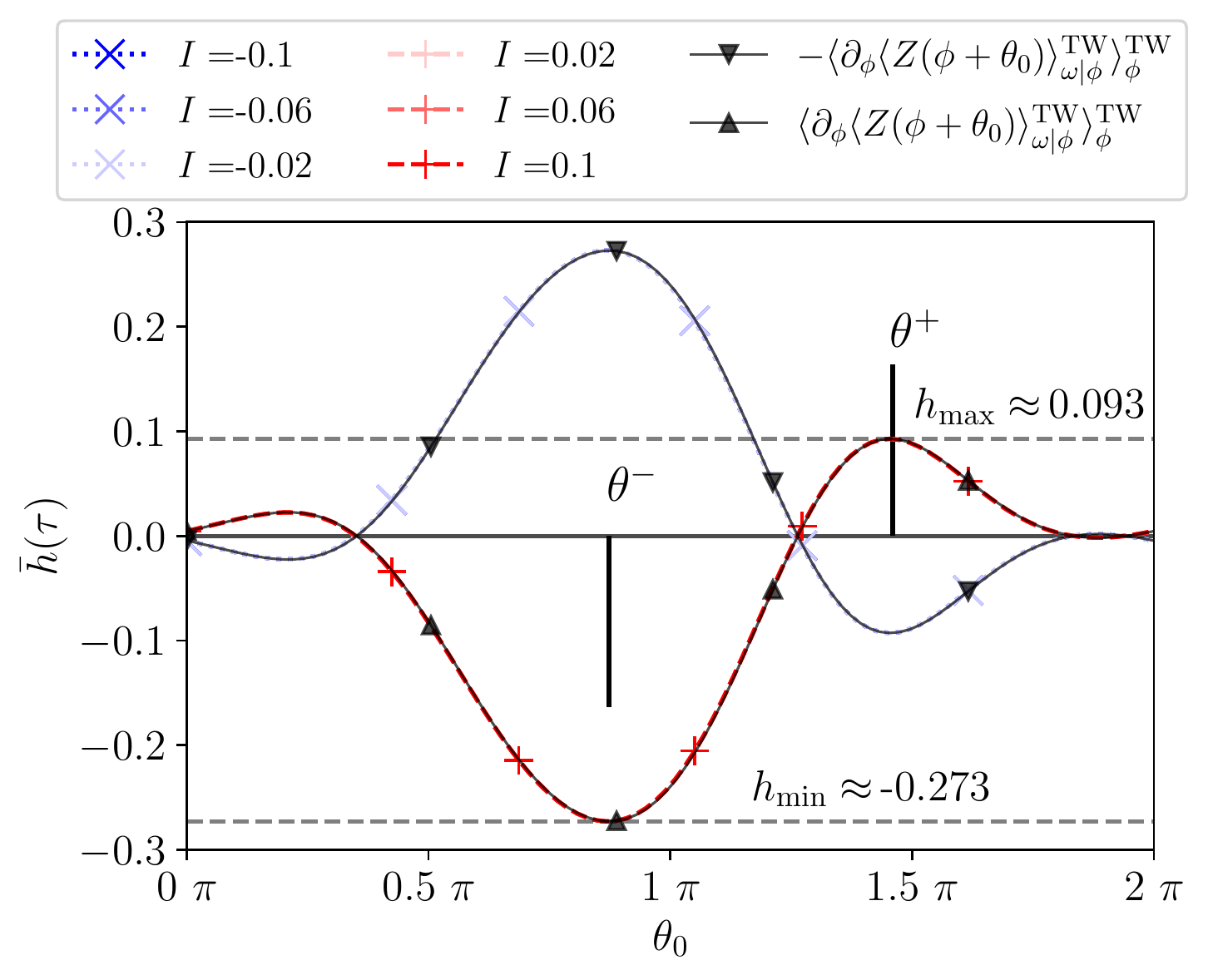}
	\caption{Charge-relative entropy step after monophasic stimulation $\bar{h}(\tau)$ as a function of the mean-field phase at the onset of stimulation $\theta_0$. Blue curves indicate a negative first pulse $I<0$, while red curves correspond to a positive first pulse $I>0$. The black solid curves with lower and upper orientated triangle markers are the first-order approximation from Eq.~(\ref{eq:H_mono_approx}). $\theta^-$ and $\theta^+$ denote the position of the minimum and maximum of the theoretical curves, see Eq.~(\ref{eq:h_min}) and Eq.~(\ref{eq:h_max}). The black bars at their positions indicate the negative and positive pulse, respectively, that lead to the maximal possible entropy change for their respective polarity.
	}
	\label{fig:H_mono}
\end{figure}

\begin{figure*}
	\includegraphics[width=0.87\textwidth]{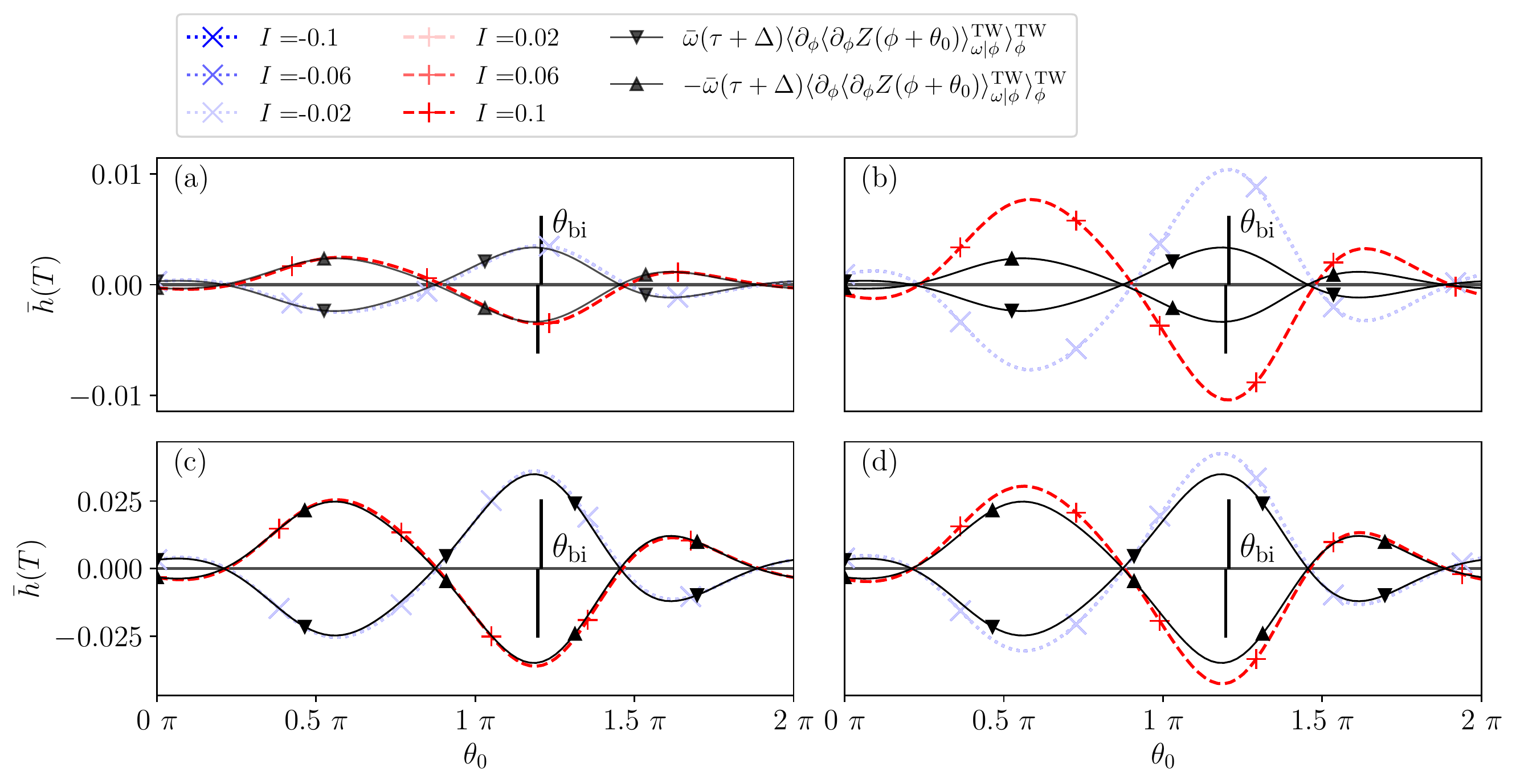}
	\caption{Charge-relative entropy step after biphasic stimulation $\bar{h}(T)$ as a function of the mean field phase angle at the onset of stimulation $\theta_0$ for four different cases of off-time $\Delta$ and asymmetry $K$: Panels (a,b,c,d) correspond to the standard symmetric gap-less  stimulus ($K=1$ and $\Delta=0$) and stimuli with $K=5$ and $\Delta=0$, $K=1$ and $\bar{\omega}\Delta \approx 0.03$, and $K=5$ and $\bar{\omega}\Delta \approx 0.03$, respectively.
	Blue curves indicate a negative first pulse $I<0$, while red curves correspond to a positive first pulse $I>0$. The black curves show the $\mathcal{O}(\tau)$ approximation from Eq.~(\ref{eq:H_bi_approx_Delta_small}) and $\theta_\text{bi}$ denotes the position of its maximum absolute value, see Eq.~(\ref{eq:theta_bi}). The black bars at $\theta_\text{bi}$ indicate the succession of a negative first pulse, followed by a second positive one, as a pulse delivered at $\theta_\text{bi}$ yields a desynchronizing effect only if $I<0$ in this example.
	}
	\label{fig:H_bi}
\end{figure*}

We perform numerical simulations of the model described in Sec.~\ref{sec:results_bimodal} using a spectral method, with $15$ Fourier modes for each subpopulation, and integrate in time using the fourth-order Runge-Kutta scheme. Before applying stimulation, we set the initial state of the system to a von Mises distribution and integrate it for a time of $T_\text{eq} = 500$ to reach the equilibrium solution. Details of the numerical technique are described in Appendix~\ref{sec:apx_methods}.

To model a TW state, we choose the parameters of coupling, diffusion, and frequency distribution as in Sec.~\ref{sec:results_bimodal}, Fig.~\ref{fig:bifurcation_diagram}(a,b): $\bar{\w} = 10$, $\eta=1$, $D=1$, $\epsilon=4$, $\alpha=0.4$. When the equilibrium state is achieved at time $T_\text{eq}$, we shift the obtained distributions $P(\phi, T_\text{eq} \,|\, \w^\mp)$ in $\phi$ to start stimulation at specific mean-field phase angles. We fix $t=0$ by the onset of the stimulation and denote the mean-field phase  at this time by $\theta(0) = \theta_0$. Having in mind possible neuroscience applications, see Sec.~\ref{sec:intro_requirements}, we relate time parameters $\tau$ and $\Delta$ to the central frequency $\bar{\omega}$, using it as an approximation for the actual oscillation frequency $\omega_0$. For all conducted numerical simulations we used the value $\tau = 0.001$ ($\bar{\w} \tau \approx 0.003 \pi$) as the duration of the first pulse.

As outlined in Sec.\ref{sec:information_entropy}, we want to use the entropy change per injected charge as the performance measure of a stimulus. This quantity $h$ is defined by Eq.~(\ref{eq:h_def}). However, depending on the sign of $I$, positive values of $h$ can correspond to both positive and negative entropy change. 
Thus, in the following, we use 
\begin{align}
    \bar{h}(t) 
    := \text{sgn}(I) h(t) 
    = \frac{H(t)-H(0)}{|I|\tau}
    \label{eq:h_bar_def}
    \,.
\end{align}
as the final performance measure. We refer to it as charge-relative entropy change. In contrast to $h$, $\bar{h}$ allows us to compare stimuli of different $I$ and $\tau$ in the sense whether they increase or decrease the synchrony level: If $\bar{h}$ is positive, the total change in entropy is positive as well. 

At first, we study the response to a monophasic stimulation as a function of $\theta_0$ for several stimulation currents $I$. The results are illustrated in Fig.~\ref{fig:H_mono}. By comparing the numerically calculated charge-relative entropy change $\bar{h}$ (blue dotted and dashed red curves) to the theoretical approximation from Eq.~(\ref{eq:H_mono_approx}) (black solid curves with triangles down and up), we see a good agreement between the theoretical and numerical curves. Since all curves for $I$ of the same sign coincide in Fig.\ref{fig:H_mono}, only one blue and one red curve are visible. Thus, for the chosen value of $\tau$, the derived first-order approximation is highly accurate.

The black bars in Fig.~\ref{fig:H_mono} indicate the stimulation onset phases $\theta^-$ and $\theta^+$ that respectively, minimize and maximize $h$ in the first-order approximation. As this approximation is highly accurate in this particular case, the monophasic stimulus is most effective for the purpose of desynchronization for a negatively charged pulse ($I<0$) delivered at $\theta^-$ and for a positively charged pulse ($I>0$) delivered at $\theta^+$. Out of the two, the first case is favorable, as $|h_\text{min}|>|h_\text{max}|$ here.   

\subsection{\label{sec:results_bi} Response to biphasic stimulation}

\begin{figure}[htb]
	\includegraphics[width=0.48\textwidth]{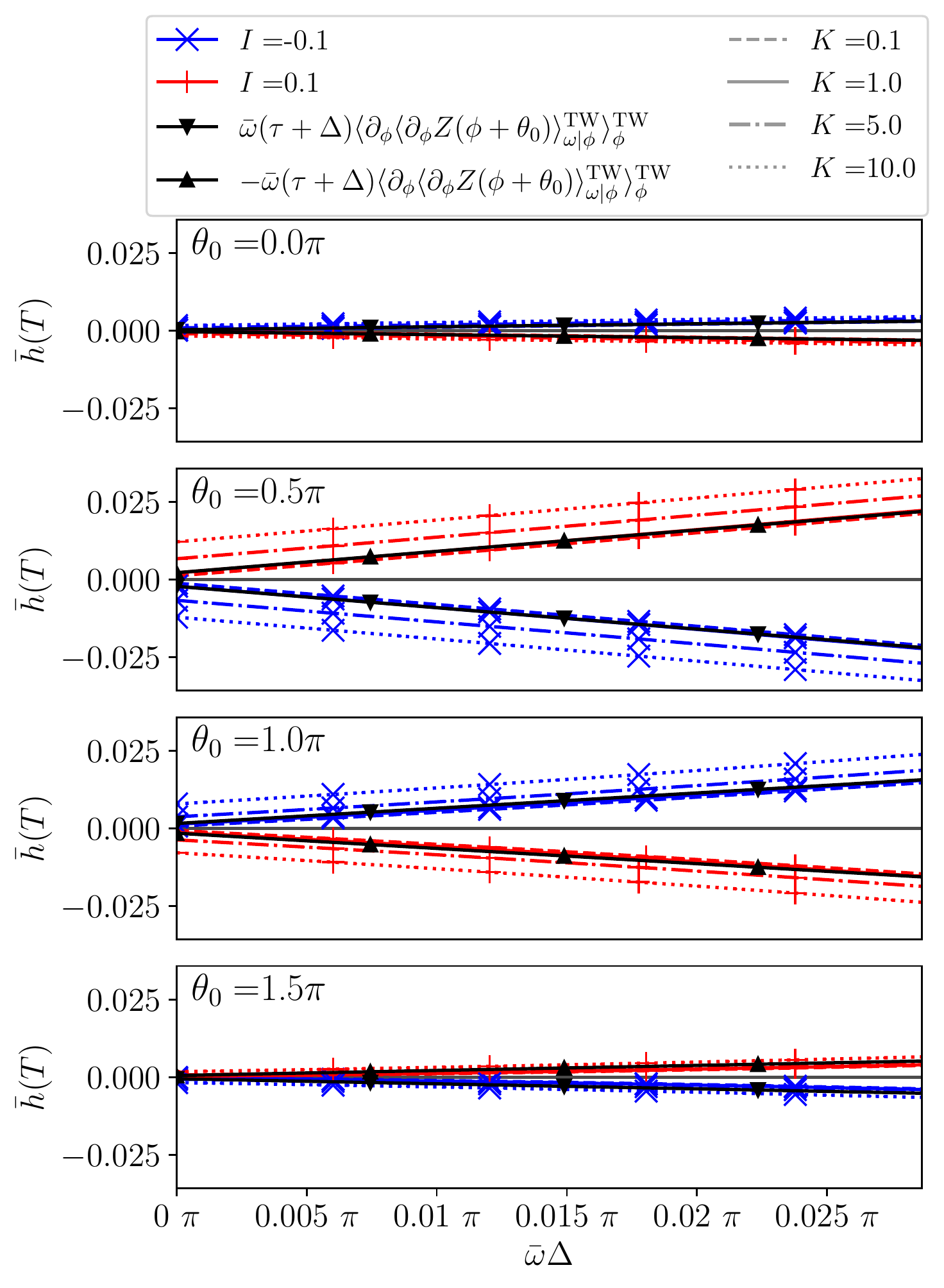}
	\caption{Charge-relative entropy step after biphasic stimulation $\bar{h}(T)$ as a function of off-time $\Delta$ in a range of $\bar{\omega} \Delta \leq 0.03 \pi$ for several mean-field phases at the onset of stimulation $\theta_0$ and asymmetry factors $K$, see legend. Markers are set to guide the eye but are not exclusive positions of data points.}
	\label{fig:H_bi_gap_small}
\end{figure}

\begin{figure}
	\includegraphics[width=0.48\textwidth]{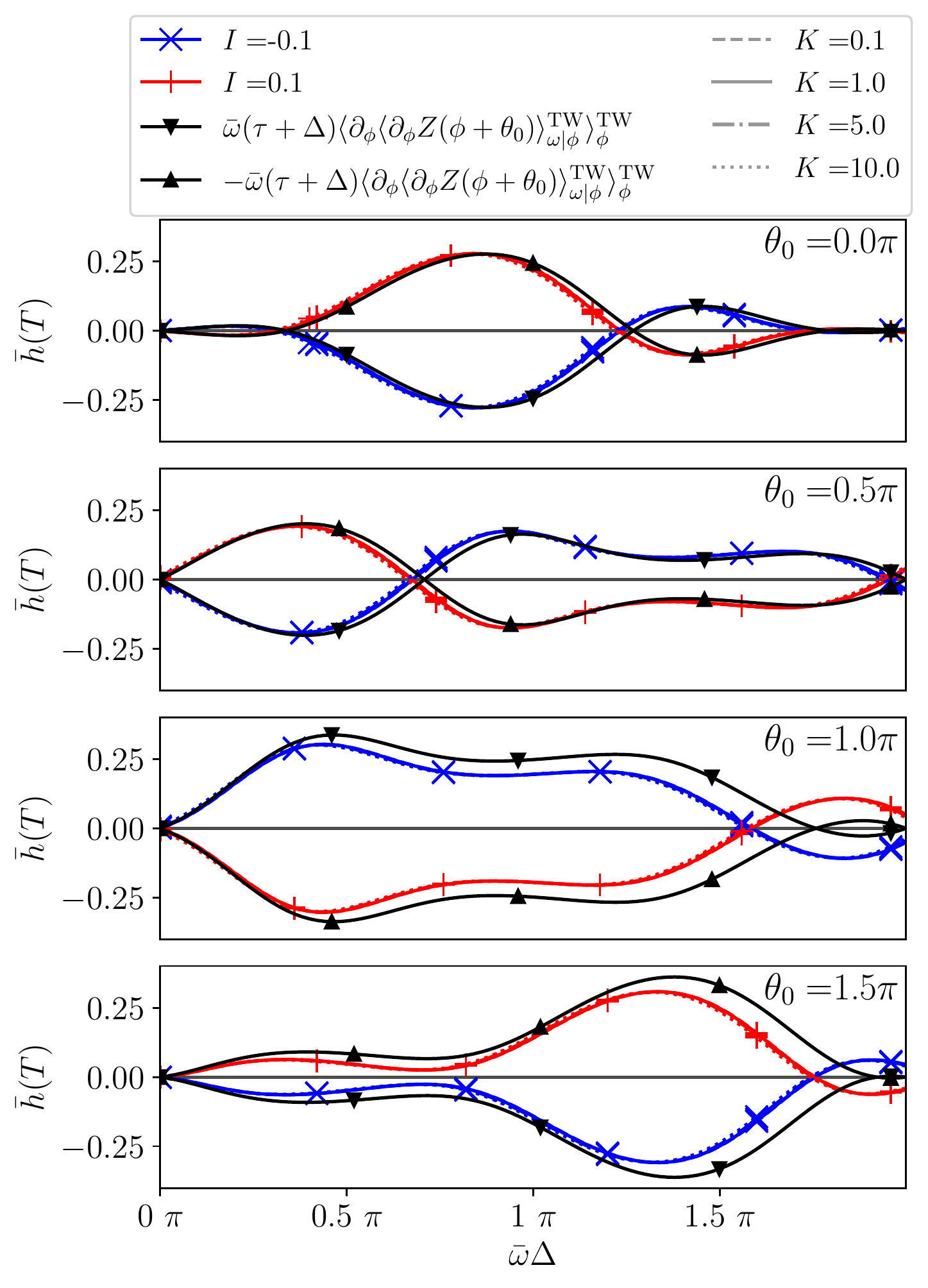}
	\caption{Charge-relative entropy step after biphasic stimulation $\bar{h}(T)$ as a function of off-time $\Delta$ in a range of $\bar{\omega} \Delta \leq 2\pi$ for several mean-field phases at the onset of stimulation $\theta_0$ and asymmetry factors $K$, see legend. Markers are set to guide the eye but are not exclusive positions of data points.}
	\label{fig:H_bi_gap_large}
\end{figure}

In this subsection, we compare the numerical results of the system's response to biphasic stimulation with the analytical predictions derived in Sec.~\ref{sec:information_entropy}. 

Starting with the case of small $\Delta$, i.e. $\w_0\Delta \ll 1$, we illustrate the charge-relative entropy step $\bar{h}$ after a biphasic pulse in Fig.~\ref{fig:H_bi} for the standard symmetric ($K=1$) gap-less ($\Delta=0$) pulse shape and its variations, namely $K=5$ and $\bar{\w}\Delta \approx 0.03 \pi$, and compare them to the derived approximation from Eq.~(\ref{eq:H_bi_approx_Delta_small}). Note that the derived approximation does not depend on $K$ and thus predicts the same change in entropy for cases (a,b) and (c,d) in Fig.~\ref{fig:H_bi}. 

As all numerical curves with the same sign in $I$ coincide, we confirm the linear dependence of the total entropy change $H(T)-H(0)$ on $I$. In the symmetric gap-less case (a), we see a good agreement between theory and the numerical simulation. For the asymmetric gap-less biphasic pulse in (b) with $K=5$, the theoretical curve does not account for the correct magnitude of $\bar{h}$ but reproduces approximately everywhere the correct sign, and $\theta_\text{bi}$ identifies the most effective state almost correctly. The maximum $\bar{h}$ in (b) is approximately twice as high as the maximum $\bar{h}$ in (a).

In both the symmetric (c) and asymmetric (d) case for a non-zero gap $\Delta$, the theoretical curves agree very well with the numerical results as both cases yield similar values of $\bar{h}$, although $K=5$ in (d). The magnitude of $\bar{h}$ is about ten times larger compared to the gap-less cases (a,d) due to the increase in off-time $\Delta$. 

In all four cases, the state for the most effective stimulation remains approximately at $\theta_0 \approx \theta_\text{bi} \approx 1.2 \pi$. It differs from the most effective state for monophasic stimulation $\theta^-$ identified from Fig.~\ref{fig:H_mono}. Also, the maximum $\bar{h}(\tau)$ from the monophasic case in Fig.~\ref{fig:H_mono} is about seven times higher than the overall maximum $\bar{h}(T)$ from all cases in Fig.~\ref{fig:H_bi}. 

Figure~\ref{fig:H_bi_gap_small} illustrates the dependence of $\bar{h}$ to off-times $\Delta$ in a small range up to $\bar{\w}\Delta \approx 0.03 \pi$. As predicted, we see a linear dependence with a slope that depends on $\theta_0$. The chosen values for the asymmetry $K$ do not affect the slope in this limit but only shift the curve. Larger asymmetries $K>1$ shift $\bar{h}$ towards larger values, whereas the curves for $K=0.1$ show even weaker responses as for $K=1$.

We now turn to the case of off-times comparable to the oscillation period, i.e. $\omega_0\Delta \approx 1$. The dependence of $\bar{h}(T)$ on off-times $\Delta$ up to $\bar{\w}\Delta = 2\pi$ is shown in Fig.~\ref{fig:H_bi_gap_large}. The numerical and theoretical curves clearly exceed the regime of linear dependence on $\Delta$. In this range, for each shown starting state $\theta_0$, one can find values of $\Delta$ for which $\bar{h}(T)$ enters the order of magnitude of the hypothetically achievable value $|h_\text{max}-h_\text{min}| \approx 0.36$ from Eq.~(\ref{eq:h_max}) and Fig.~\ref{fig:H_mono}. Here, all investigated asymmetry parameters $K$ show no significant effect on the achieved response. The theoretical predictions (solid black curves) show a fair agreement with the numerical results, especially for $\theta_0 \in \{0, ~0.5\pi\}$ and values of $\omega_0 \Delta \leq \pi$, but deviate strongly for large values of $\Delta$ as relaxation processes become prominent.  

\begin{figure}
	\includegraphics[width=0.48\textwidth]{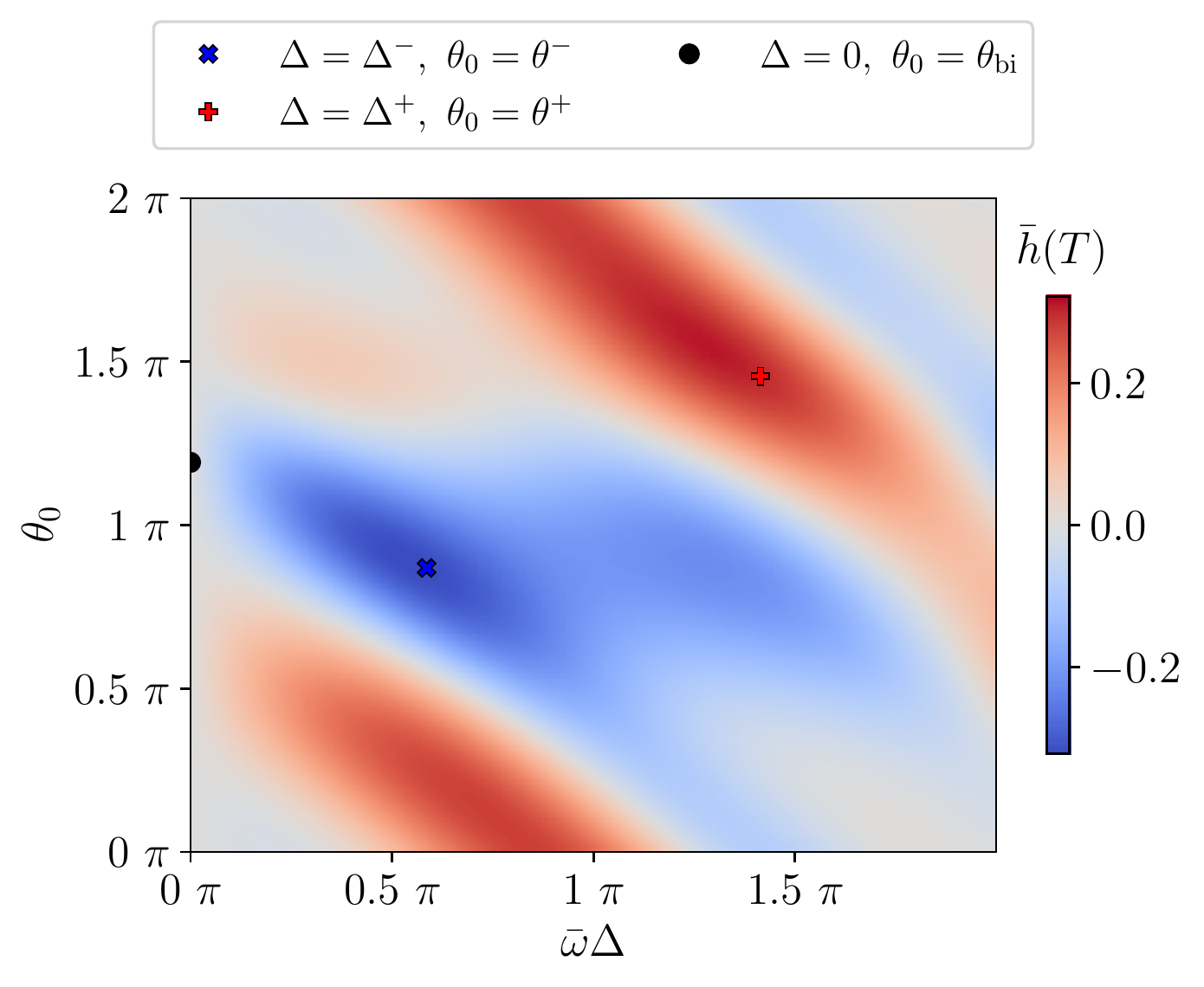}
	\caption{Heat map of charge-relative entropy step after biphasic stimulation $\bar{h}(T)$ in the parameter region of off-time $\Delta$ and starting state $\theta_0$ shown for a pulse of $I=0.1$ and asymmetry factor $K=1$. For $I<0$ the color-coding is inverted. The symbols represent the theoretical predictions for the most effective starting states and corresponding off-times: the black circle as $\theta_\text{bi}$ in the limit $\Delta = 0$, the blue cross as $\theta^-$ with $\Delta^- := \frac{(\theta^+ - \theta^-) \mod 2\pi}{\bar{\w}} $, and the red plus as $\theta^+$ with $\Delta^+ := \frac{(\theta^- - \theta^+) \mod 2\pi}{\bar{\w}}$.}
	\label{fig:H_bi_heatmap_theta_gap}
\end{figure}

In order to validate the optimality of the designed large-$\Delta$ stimulation scheme from Sec.~\ref{sec:information_entropy}, we plot the biphasic response $\bar{h}(T)$ vs. $\theta_0$ and $\Delta$ as a heat map of both parameters in Fig.~\ref{fig:H_bi_heatmap_theta_gap}. The maximal achieved $\bar{h}(T)$ are of the same order of magnitude as $h_\text{max}$ and $h_\text{min}$ and thus about $100$ times larger than the response without any off-time $\Delta=0$ (cf. Fig.~\ref{fig:H_bi}a) and around $10$ times larger than the response for small $\Delta$ (Fig.~\ref{fig:H_bi}c,d). Indeed, the starting states $\theta^\mp$ with off-times $\Delta^\mp$ adapted to their spatial distance are inside the parameter regions of $\Delta$ and $\theta_0$, where $\bar{h}(T)$ attains the largest values. Thus, designing an effective biphasic stimulation scheme based on where a monophasic pulse is most effective seems promising.

\begin{figure}
	\includegraphics[width=0.48\textwidth]{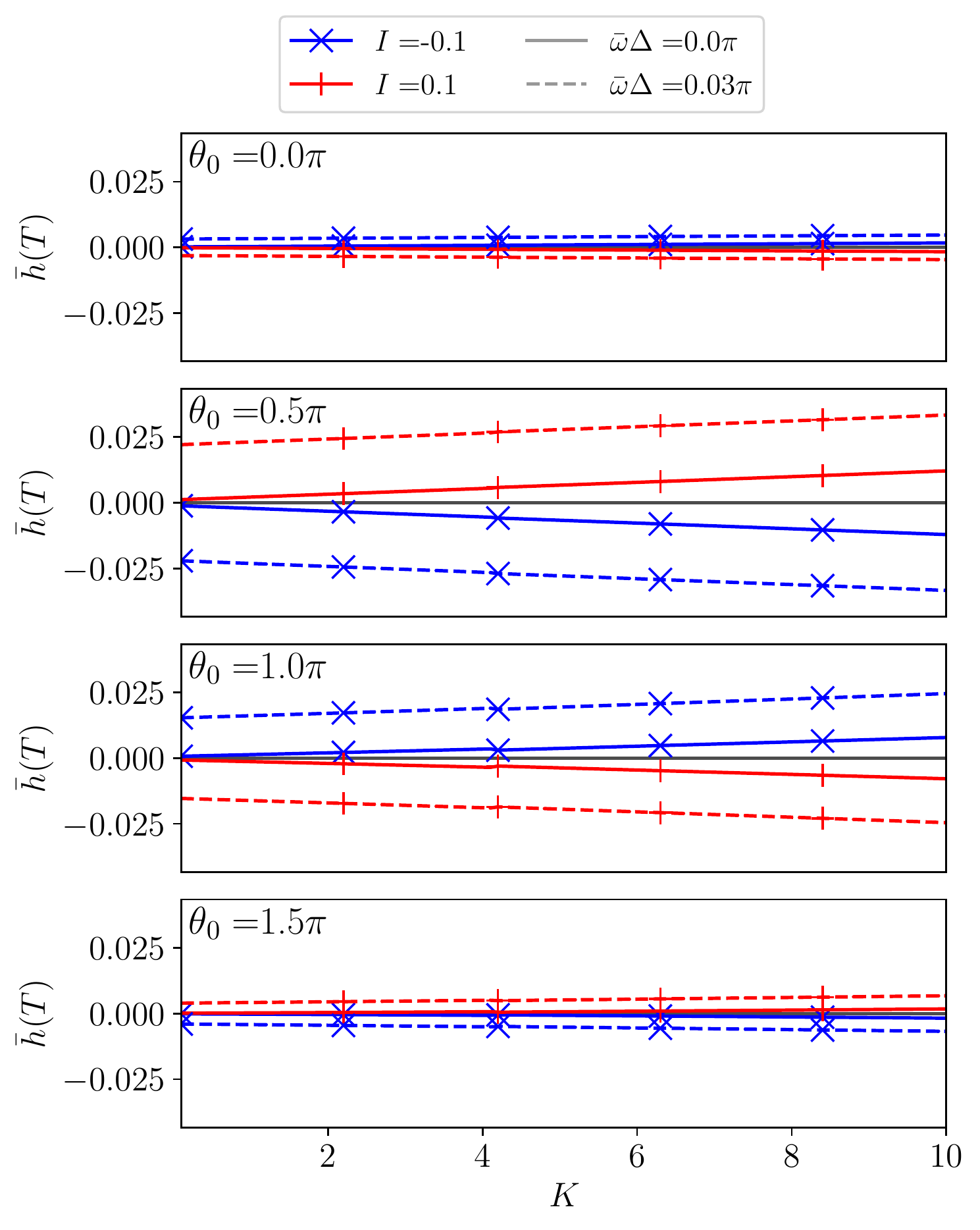}
	\caption{Charge-relative entropy step after biphasic stimulation $\bar{h}(T)$ as a function of asymmetry parameter $K$ in the range $0.1 \leq K \leq 10$ for several mean field phase angles at the onset of stimulation $\theta_0$ and off-times $\Delta$, see legend. Markers are set to guide the eye but are not exclusive positions of data points.}
	\label{fig:H_bi_K}
\end{figure}

Finally, we investigate the dependence of $\bar{h}(T)$ on the asymmetry parameter $K$ in Fig.~\ref{fig:H_bi_K}. For the two displayed curves, for $\Delta = 0$ and $\bar{\w}\Delta = 0.03 \pi$, the dependence on $K$ follows a linear trend, whose slope depends on the starting state $\theta_0$. Here, larger $K$ leads to a stronger system response in both synchronizing and desynchronizing directions, whereas $K<1$ leads to an even weaker response than $K=1$. To summarize, in the limit of small $\Delta$, an asymmetry parameter $K>1$ can lead to significantly larger system response. The dependence on $K$ vanishes almost entirely for high values of $\Delta$ where $\bar{\w}\Delta \approx 1$, see Fig.~\ref{fig:H_bi_gap_large}.

To conclude this section, we highlight that for the analytical expressions of the $\mathcal{O}(\tau)$ approximations used in the last two subsections, only the function 
\begin{align}
    \theta \mapsto \langle \partial_\phi \langle Z(\phi + \theta) \rangle^\text{\tiny TW}_{\w|\phi} \rangle^\text{\tiny TW}_\phi
\end{align}
is needed. In particular, it also suffices for the predictions of the biphasic pulse for small $\Delta$, as its derivative with respect to $\theta$ is exactly the term used for the small-$\Delta$-approximation:
\begin{align}
    \partial_\theta \langle \partial_\phi \langle Z(\phi + \theta) \rangle^\text{\tiny TW}_{\w|\phi} \rangle^\text{\tiny TW}_\phi
    = 
    \langle \partial_\phi \langle \partial_\phi Z(\phi + \theta) \rangle^\text{\tiny TW}_{\w|\phi} \rangle^\text{\tiny TW}_\phi
    \,.
\end{align}

\section{\label{sec:discussion} Discussion and conclusion}

In summary, we used a noisy Kuramoto-Winfree model to explore the effect of charge-balanced, biphasic stimulation on the degree of synchrony in a population of non-identical oscillators with frequency-dependent phase sensitivity functions.  We derived general formulas which quantify the collective response in terms of the phase distribution's information entropy. These formulas allow predicting the most effective state, i.e., the mean-field phase, for the stimulation's application, both in the limit of short off-stimulation times and in the case of off-stimulation times comparable to the oscillation period. We found a relationship between this most effective state and the first and second derivatives of the phase response curves. We verified our general approach exploiting a tractable two-frequency ensemble model with a particular choice of corresponding phase response curves and demonstrated a good agreement with the theory and numerical simulations.

Whether it enhances or suppresses synchrony, the effect of a pulse depends on the mean-field phase $\theta_0$ when the stimulation is applied and on the pulse's polarity $I$. This fact calls for a feedback-based stimulation controller that monitors the system's state and adjusts the stimulation polarity accordingly.

Our research aimed to analyze the dependence of the biphasic stimulation's efficiency on the off-stimulation time $\Delta$. We confirmed that variation of $\Delta$ could increase the system's response by orders of magnitude. However, this effect strongly depends on the allowed range of $\Delta$. The challenge in designing effective stimulation schemes for small $\Delta$ is to minimize the canceling impact of the opposite pulses. Thus, the second derivative of the conditional frequency-averaged PRC determines the most sensitive initial state for biphasic stimulation in the limit of small $\Delta$. Suppose now that $\Delta$ is allowed to be comparable to the oscillation period. In that case, it is most effective to apply the negative and positive monophasic stimuli at minimum and maximum of $h(t)$, respectively. Such biphasic stimulus provides maximal cumulative impact on the entropy. Thus, depending on whether small or large values of $\Delta$ are allowed in a particular application, we will use different strategies to optimize the stimulus's shape and find the most sensitive state.

We emphasize two general properties of our results derived in Sec.~\ref{sec:model}. First, the sensitivity of the oscillators to stimulation, i.e., PRC,  can be determined by an arbitrary parameter, not necessarily by frequency. In this case, the conditional frequency average shall be substituted by the conditional averaging over this parameter. Second, the validity of our results goes beyond the model of sine-coupled oscillators. Since our central assumption is the existence of the TW state, the exact form of the inter-oscillator coupling resulting in this state is unimportant. We expect our results to hold for any population that can be described in the phase approximation.

As a direction of further studies, we mention the extension of the developed theory to cover more complex autonomous states than a traveling wave state.

\begin{acknowledgments}
We acknowledge useful discussions with A. Pikovsky. The work was supported by Deutsche Forschungsgemeinschaft (DFG, German Research Foundation) – Project-ID 424778381 – TRR 295.
\end{acknowledgments}

\section*{Data availability}
Data sharing is not applicable to this article as no new data were created or analyzed in this study.

\appendix

\section{\label{sec:apx_stat_averages} Notations for averaged quantities}
In this section, we summarize the notations for averaging. The notation $\langle \cdot \rangle$ without a subscript, used only in Sec.~\ref{sec:model_1}, denotes the average over the realizations of the white noise $\xi$. We shall distinguish it from similar notation with different subscripts used in Sec.~\ref{sec:model_macroresponse}.

Let $f(\phi,\w)$ be an arbitrary function of the random variables $\phi$ and $\omega$. (Below, we omit the time dependence  of $f$ and distributions $P$, $\rho$ for brevity.)
We can average $f(\phi,\w)$  over the phase or frequency to obtain, respectively,
\begin{align}
    \langle f\rangle_\phi &= \int_0^{2\pi} f(\phi, \w) \rho(\phi) \dd\phi
    \label{eq:stat_average_phase}
\end{align}
and
\begin{align}
    \langle f\rangle_\w &= \int_{-\infty}^\infty f(\phi, \w) g(\w) \dd\w
    \label{eq:stat_average_freq}
    \,.
\end{align}
Since we can interchange the order of integration, these two averages commute: 
\begin{align}
    \langle \langle f \rangle_\w \rangle_\phi
    = \langle \langle f \rangle_\phi \rangle_\omega 
    \,.
\end{align}
Next, with the help of conditional phase distribution $P(\phi \,|\, \omega)$ we define the \textit{conditional phase average}
\begin{align}
    \langle f\rangle_{\phi|\w} &= \int_0^{2\pi} f(\phi, \w) P(\phi \,|\, \w) \dd\phi\;,
    \label{eq:stat_average_phase_cond}
\end{align}
that is a function of $\w$. Likewise, using the conditional frequency distribution $P(\omega \,|\, \phi)$ we compute the \textit{conditional frequency average}
\begin{align}
    \langle f\rangle_{\w|\phi} &= \int_{-\infty}^\infty f(\phi, \w) P(\w \,|\, \phi) \dd\w
    \label{eq:stat_average_freq_cond} 
\end{align}
as a function of $\phi$. The \textit{global average} over both random variables is defined by
\begin{align}
    \langle f\rangle_{\phi,\omega} 
    &= \int_0^{2\pi} \int_{-\infty}^\infty f(\phi, \w) P(\phi, \w) \dd\w \dd\phi
    \,.
\end{align}
Recalling the definition of conditional probability densities
\begin{align}
    P(\phi, \omega) = P(\phi \,|\, \w)g(\w) =  P(\w \,|\, \phi)\rho(\phi) \;,
\end{align}
we express the global average as
\begin{align}
    \langle f\rangle_{\phi,\omega} 
    = \langle \langle f\rangle_{\phi|\w} \rangle_\w
    = \langle \langle f\rangle_{\w|\phi} \rangle_\phi
    \,.
\end{align}
Note that generally, $\w$ and $\phi$ are not independent random variables and thus $\langle \langle f \rangle_\w \rangle_\phi \neq \langle f\rangle_{\phi,\omega}$. The consecutive averages and the global average are equal only if $\phi$ and $\omega$ are independent.

\begin{widetext}
\section{\label{sec:apx_Hder} Time derivative of the information entropy}
In order to estimate the total change in information entropy $H$ for short time scales, we employ the time derivative $\dot{H}$. Here, we outline its derivation.
Starting with its definition in Eq.~(\ref{eq:H_def}), the time derivative of the information entropy is given in general by 
\begin{align}
    \dot{H}(t) = - \int_0^{2\pi} \ln(\rho) \partial_t \rho\dd\phi
    \label{eq:dH_general}
    \;. 
\end{align}
Since the global phase distribution $\rho(\phi)$ is the frequency average of the conditional phase distribution $P(\phi \,|\, \w)$ (see Eq.~(\ref{eq:def_rho})), we can insert the $\w$-average of Eq.~(\ref{eq:Kuramoto_PDE}) for $\partial_t \rho$, which yields
\[
\dot{H}(t) = - \int_0^{2\pi} \int_{-\infty}^\infty \ln(\rho) g(\w) \left\{-\partial_\phi \left[(\w + \e R \sin(\theta-\phi) + J(t) Z(\phi;\w))P(\phi, t \,|\, \w)\right ] + D\partial^2_\phi P(\phi, t \,|\, \w) \right\} \dd\w \dd\phi\;.
\]
Since boundary terms vanish due to the $2\pi$-periodicity in $\phi$ in all functions involved, integration by parts gives
\[
\dot{H}(t) = \int_0^{2\pi} \int_{-\infty}^\infty \frac{\partial_\phi \rho}{\rho} g(\w) \left\{-[\w + \e R \sin(\theta-\phi) + J(t) Z(\phi;\w)]P(\phi, t \,|\, \w) + D\partial_\phi P(\phi, t \,|\, \w) \right\} \dd\w \dd\phi\;,
\]
which can be rewritten as
\[
\dot{H}(t) =\int_0^{2\pi} \partial_\phi \rho 
	\left[ 
	    - \int_{-\infty}^\infty \w \frac{P(\phi, t \,|\, \w)g(\w)}{\rho(\phi,t)} \dd\w
	    -\e R \sin(\theta-\phi)
	    -J(t) \int_{-\infty}^\infty Z(\phi;\w) \frac{P(\phi, t \,|\, \w)g(\w)}{\rho(\phi,t)} \dd\w
	    + D  \frac{\partial_\phi \rho}{\rho}
	\right] \dd\phi \;.
\]
Using the conditional frequency average Eq.~(\ref{eq:stat_average_freq_cond}) and another integration by parts, for all terms except the diffusion term we obtain 
\[
\dot{H}(t) = \int_0^{2\pi} \rho 
	\left[ 
	    \partial_\phi \langle \w \rangle_{\omega|\phi}
	    - \e R \cos(\theta-\phi)
	    + J(t) \partial_\phi \langle Z \rangle_{\omega|\phi}
	\right] \dd\phi 
	 + D \int_0^{2\pi} \frac{(\partial_\phi \rho)^2}{\rho} \dd\phi \;.
\]
Recalling the phase-average Eq.~(\ref{eq:stat_average_phase}) and the definition of the mean field in Eq.~(\ref{eq:mfield}) we finally write
\begin{equation}
  \dot{H}(t) =  \langle  \partial_\phi \langle \w \rangle_{\omega|\phi} \rangle_\phi
	    - \e R^2 
	    + J(t) \langle \partial_\phi \langle Z \rangle_{\omega|\phi} \rangle_\phi
	 + D \int_0^{2\pi} \frac{(\partial_\phi \rho)^2}{\rho} \dd\phi \;.
	\label{eq:dH_Kuramoto}
\end{equation}
As outlined in Section~\ref{sec:model_macroresponse}, we assume the system to be in a traveling wave (TW) state before the stimulation starts. Due to the short time scales of the stimulation with respect to all involved relaxation processes, we approximate the time evolution of $\rho$ by the PDE~(\ref{eq:TW_PDE}). If Eq.~(\ref{eq:TW_PDE}) is inserted into Eq.~(\ref{eq:dH_general}) instead of the exact PDE~(\ref{eq:Kuramoto_PDE}),  then the rotation term $\omega_0$ is constant for all subpopulations and thus the partial derivative with respect to $\phi$ of $\langle \w_0 \rangle_{\w|\phi} = \w_0$ vanishes.

Hence, the only term that remains from the general expression Eq.~(\ref{eq:dH_Kuramoto}) is the stimulation term:
\begin{align}
	\dot{H}(t) = J(t) \langle \partial_\phi \langle Z \rangle_{\omega|\phi} \rangle_\phi
	\label{eq:dH_TW_stim}
	\;.
\end{align}
With the assumption, that the stimulation $J$ does not change the shape of $\rho$ significantly within the time scale of its duration, we can further approximate $\dot{H}$ using the static averages with the TW solutions $P_\text{\tiny TW}(\phi \,|\, \w)$ from Eq.~(\ref{eq:average_TW}):
\begin{align}
	\dot{H}(t) \approx J(t) \langle \partial_\phi \langle Z(\phi + \theta(t)) \rangle^\text{\tiny TW}_{\omega|\phi} \rangle^\text{\tiny TW}_\phi
	\;.
\end{align}
This result is exact for the start of the stimulation at $t=0$ but it is approximate for $t>0$.

\section{\label{sec:apx_methods} Methods}

For numerical simulations of the bimodal noisy Kuramoto model described in Sec.~\ref{sec:results_bimodal}, we employ a spectral method~\cite{press_numerical_2007}: Exploiting the $2\pi$-periodicity in $\phi$ we write both conditional phase distributions $P(\phi,t \,|\, \w^\mp)$ and the PRCs $Z^\mp$ in terms of their complex Fourier modes $P^\mp_n(t)$ and $Z^\mp_n$, respectively:
\begin{align}
    P^\mp_n(t) 
    = \int_0^{2\pi} \ee^{\ii n\phi} P(\phi,t \,|\, \w^\mp) \dd\phi 
    \quad \text{and} \quad
    Z^\mp_n = \frac{1}{2\pi} \int_0^{2\pi} \ee^{\ii n \phi} Z(\phi;\w^\mp) \dd\phi
    \qquad n \in \mathbb{Z}
    \,.
\end{align}
In this way we transform Eq.~(\ref{eq:Kuramoto_PDE}) into an infinite system of coupled ordinary differential equations (ODEs) for both subpopulations:
\begin{align}
    \dot{P}^\mp_n = 
     \ii n(\bar{\omega} \mp \eta) P^\mp_n
    + \frac{\epsilon n}{2}(R\ee^{\ii \theta} P^\mp_{n-1} - R\ee^{-\ii \theta} P^\mp_{n+1})
    + \ii n J(t) \sum_{k\in \mathbb{Z}} Z^\mp_{n-k} P^\mp_k
    - n^2 D P^\mp_n
    \qquad n \in \mathbb{Z}
    \,.
\end{align}
The mean field is given by 
\begin{align}
    R\ee^{\ii \theta} = \alpha P^-_1 + (1-\alpha) P^+_1 
    \,.
\end{align}
Since the conditional distributions are real functions their modes obey $P_{-n} = P^*_{n}$. Also, due to normalization, $P^\mp_0 = 1$ for all times. Thus, to numerically simulate this system we truncate the Fourier expansion at a mode number $N_P=15$ and set all higher modes to $0$. Thus, we have to integrate $N_P$ coupled ODEs for both subpopulations.

The inverse transformation from the Fourier modes to the function in physical space is given by
\begin{align}
    P(\phi,t \,|\, \w^\mp) 
    = \frac{1}{2\pi} \sum_{n \in \mathbb{Z}} P^\mp_n \ee^{-\ii n\phi}
    \approx  \frac{1}{2\pi} + \frac{1}{\pi} \sum_{n=1}^{N_P} \Re(P^\mp_n \ee^{-\ii n\phi}) 
    \label{eq:P_from_modes}
    \,.
\end{align}
Similarly, we represent $Z^\mp$ in terms of a finite number $N_Z$ of Fourier modes, with $Z_{-n} = Z^*_{n}$ due to reality of $Z$. To numerically represent $Z^\mp$ its Fourier mode representation is truncated at $N_Z$. 
The inverse transformation to the function in $\phi$ reads
\begin{align}
    Z^\mp(\phi) 
    = \sum_{n \in \mathbb{Z}} Z^\mp_n \ee^{-\ii n\phi}
    \approx  Z_0 + 2 \sum_{n=1}^{N_Z} \Re(Z^\mp_n \ee^{-\ii n\phi})
    \label{eq:Z_from_modes}
    \,.
\end{align}
Due to the truncation of mode numbers, the convolution term is approximated by
\begin{align}
    \sum_{k\in \mathbb{Z}} Z_{n-k} P_k
	\approx 
	 \sum_{k=1}^{\min(N_Z-n,N_P)} Z_{n+k}P^*_k
	+ \sum_{k=\max(0,n-N_Z)}^{\min(N_P,n)} Z_{n-k}P_k
	+ \sum_{k=n+1}^{\min(N_P,N_Z+n)} Z^*_{k-n} P_k
	\,,
\end{align}
where we omit the superscript $\mp$ for clarity. We integrate this system of $2N_P$ coupled ODEs in time using a fourth-order Runge-Kutta scheme with an integration step $0.0001$.

As the initial state for numerical simulations we choose for both subpopulations the same von Mises distribution 
\begin{align}
    P_\text{ini}(\phi \,|\, \w^\pm) =  \frac{\exp{\left (\frac{\epsilon R}{D}\cos\phi\right)}}{2\pi I_0(\frac{\epsilon R}{D})}
\end{align}
where $I_n$ are the modified Bessel functions of the first kind \cite{abramowitz2013}. $R$ is determined by the self-consistency condition
\begin{align}
    R I_0\left(\frac{\epsilon R}{D}\right) 
    = I_1\left(\frac{\epsilon R}{D}\right)
\end{align}
that can be equivalently formulated using the recursion relation of $I_n$~\cite{abramowitz2013} by
\begin{align}
	\left(1- \frac{2D}{\epsilon} \right) I_0 \left(\frac{\epsilon R}{D}\right) 
	= I_2\left( \frac{\epsilon R}{D}\right)
	\,,
\end{align}
which can be favorable to solve for $R$ numerically.

$P_\text{ini}$ is the TW solution of the Kuramoto model with Gaussian white noise for a unimodal discrete frequency distribution $g(\omega) = \delta(\omega - \bar{\omega})$. As the unimodal distribution is the limit case for $\eta \rightarrow0$ of Eq.~(\ref{eq:g_bimodal}), we expect for small $\frac{\eta}{\bar{\w}}$ the bimodal TW solution to be close to the von Mises distribution $P_\text{ini}$ and thus taking less time $T_\text{eq}$ to reach the equilibrium.

The entropy in Eq.~(\ref{eq:H_def}) is calculated by approximating the conditional phase distributions from its Fourier modes by Eq.~(\ref{eq:P_from_modes}), calculating the global phase distribution $\rho$ by Eq.~(\ref{eq:global_phase_distr_bimodal}), and using the numerical trapezoidal integration scheme of the Python package NumPy \cite{harris_array_2020} with $100$ equally spaced values of the function.

All figures in this article were made with the Python package Matplotlib\cite{hunter2007}. 

\end{widetext}

\nocite{*}
%



\end{document}